\documentclass[%
reprint,
superscriptaddress,
nofootinbib,
amsmath,
amssymb,
aps,
prx,
floatfix,
longbibliography
]{revtex4-2}

\usepackage{amsmath}
\usepackage{amssymb}
\usepackage{amsfonts}
\usepackage{latexsym,epsfig,bbm}
\usepackage{algorithm}
\usepackage{algorithmic}
\usepackage{soul}
\usepackage{graphicx}
\usepackage{dcolumn}
\usepackage{bm}
\usepackage{subfigure}
\usepackage{dsfont} 

\usepackage[colorlinks=false,citecolor=blue]{hyperref}
\usepackage[margin=1in]{geometry}

\usepackage{xcolor}

\renewcommand{\vec}[1]{\boldsymbol{#1}}

\newcommand{\kB}{\mbox{$k_\mathrm{B}$}}
\newcommand{\ms}{\mbox{$m^*$}}

\newcommand{\muH}{\mbox{$\mu_\mathrm{H}$}}
\newcommand{\neff}{\mbox{$n_{\mathrm{eff}}$}}
\newcommand{\nH}{\mbox{$n_{\mathrm{H}}$}}
\newcommand{\nloc}{\mbox{$n_{\mathrm{loc}}$}}
\newcommand{\nrho}{\mbox{$n_\rho$}}
\newcommand{\pp}{\mbox{$\mathfrak{p}$}}
\newcommand{\Rgs}{\mbox{$R_{\mathrm{gs}}$}}
\newcommand{\RH}{\mbox{$R_{\mathrm{H}}$}}
\newcommand{\rhoS}{\mbox{$\rho_{\mathrm{S}}$}}
\newcommand{\Rm}{\mbox{$R_{\mathrm{m}}$}}
\newcommand{\Rs}{\mbox{$R_{\mathrm{s}}$}}
\newcommand{\sigdc}{\mbox{$\sigma_\mathrm{dc}$}}
\newcommand{\Tc}{\mbox{$T_\mathrm{c}$}}

\newcommand{\Ts}{\mbox{$T^*$}}
\newcommand{\Tss}{\mbox{$T^{**}$}}

\begin{document}

\title{Characterization of two electronic subsystems in cuprates through optical conductivity}
\author{C. M. N. Kumar}
\affiliation{Institute of Solid State Physics, TU Wien, 1040 Vienna, Austria}
\affiliation{Institute of Physics, Bijeni\v{c}ka cesta 46, HR-10000, Zagreb, Croatia}
\affiliation{The Henryk Niewodniczanski Institute of Nuclear Physics, Polish Academy of Sciences, 152 Radzikowskiego Str., 31-342, Krakow, Poland}
\author{A. Akrap}
\affiliation{Department of Physics, University of Fribourg, Chemin du Mus\'{e}e 3, CH-1700 Fribourg, Switzerland}
\author{C. C. Homes}
\affiliation{Condensed Matter Physics and Materials Science Department, Brookhaven National Laboratory, Upton, New York 11973, USA}
\affiliation{National Synchrotron Light Source II, Brookhaven National Laboratory, Upton, New York 11973, USA}
\author{E. Martino}
\affiliation{Institute of Physics, \'{E}cole Polytechnique F\'{e}d\'{e}ral de Lausanne (EPFL), CH-1015, Lausanne, Switzerland}
\author{B. Klebel-Knobloch}
\affiliation{Institute of Solid State Physics, TU Wien, 1040 Vienna, Austria}
\author{W. Tabis}
\affiliation{Institute of Solid State Physics, TU Wien, 1040 Vienna, Austria}
\affiliation{AGH University of Science and Technology, Faculty of Physics and Applied Computer Science, 30-059 Krakow, Poland}
\author{O. S. Bari\v{s}i\'{c}}
\affiliation{Institute of Physics, Bijeni\v{c}ka cesta 46, HR-10000, Zagreb, Croatia}
\author{D. K. Sunko}
\affiliation{Department of Physics, Faculty of Science, University of Zagreb, Bijeni\v{c}ka cesta 32, HR-10000, Zagreb, Croatia}
\author{N. Bari\v{s}i\'{c}}
\email{nbarisic@phy.hr}
\affiliation{Institute of Solid State Physics, TU Wien, 1040 Vienna, Austria}
\affiliation{Department of Physics, Faculty of Science, University of Zagreb, Bijeni\v{c}ka cesta 32, HR-10000, Zagreb, Croatia}

\date{\today}

\begin{abstract}
Understanding the physical properties of unconventional superconductors as well as of other correlated materials presents a formidable challenge. Their unusual evolution with doping, frequency, and temperature has frequently led to non-Fermi-liquid (non-FL) interpretations. Optical conductivity is a major challenge in this context. Here, the optical spectra of two archetypal cuprates, underdoped HgBa$_2$CuO$_{4+\delta }$ and optimally-doped Bi$_2$Sr$_2$CaCu$_2$O$_{8+\delta }$, are interpreted based on the standard Fermi liquid (FL) paradigm. At both dopings, perfect frequency-temperature FL scaling is found to be modified by the presence of a second, gapped electronic subsystem. This non-FL component emerges as a well-defined mid-infrared spectral feature after the FL contribution---determined independently by transport---is subtracted. Temperature, frequency, and doping evolution of the MIR feature identify a gapped rather than dissipative response. In contrast, the dissipative response is found to be relevant for pnictides and ruthenates. Such an unbiased FL/non-FL separation is extended across the cuprate phase diagram, capturing all the key features of the normal state and providing a natural explanation why the superfluid density is attenuated on the overdoped side. Thus, we obtain a unified interpretation of optical responses and transport measurements in all analyzed physical regimes and all analyzed compounds. 
\end{abstract}

\maketitle


\section{Introduction}
\label{sec:introduction}

\begin{figure*}[!htb]
    \centering
    \includegraphics[scale=0.95]{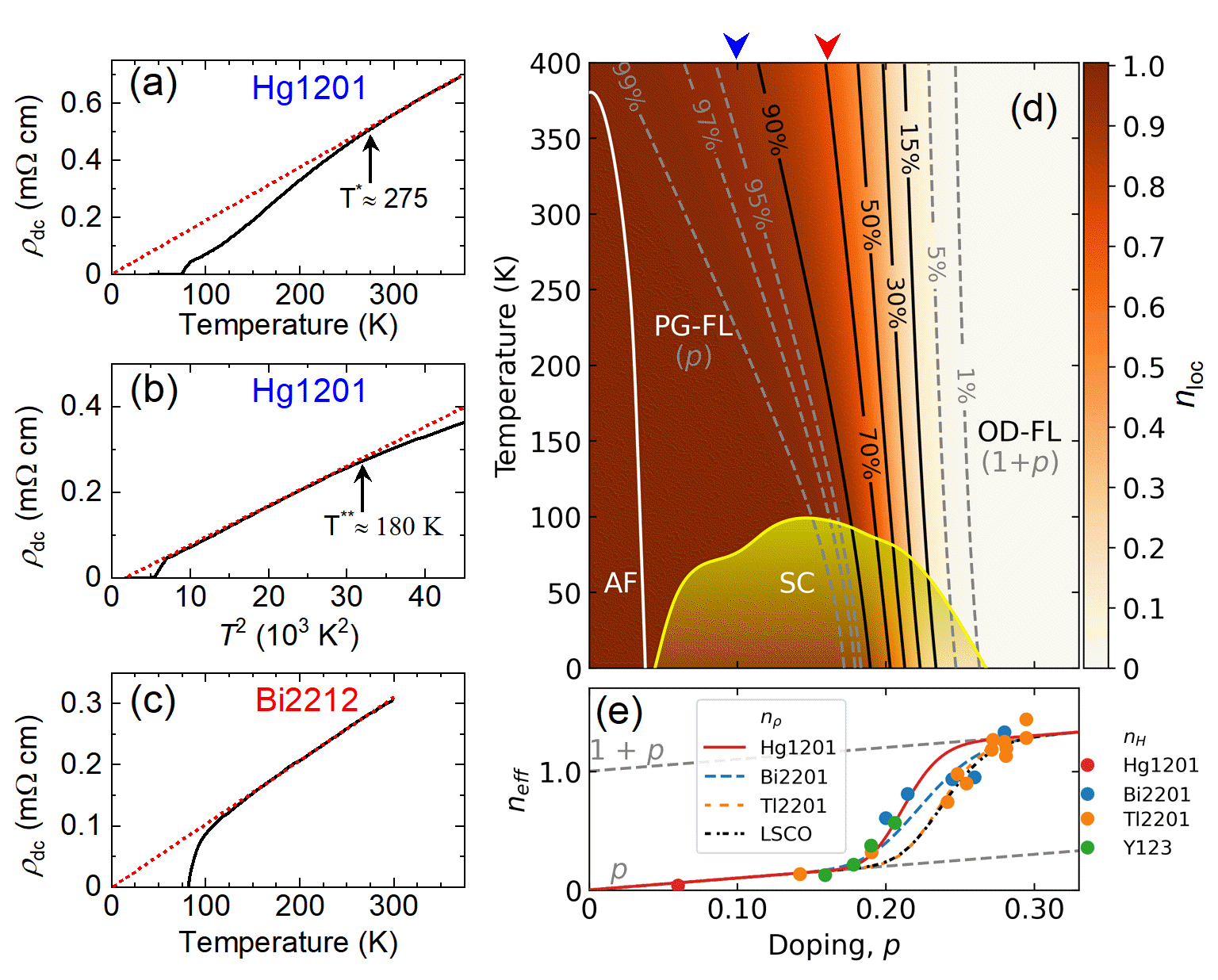}
    \caption{The temperature-dependence of the resistivity of Hg1201 at $p\approx0.1$ is (a) linear-like above $\Ts\approx 275$~K (PG temperature) and (b) purely quadratic below $\Tss\approx 180$~K, in agreement with previously published results~\cite{NBarisic13}. (c) At optimal doping ($p\approx0.16$) resistivity in Bi2212~\cite{Kendziora92} is essentially linear down to \Tc. (d) Schematic phase diagram of cuprates based on the doping and temperature dependence of the localized carrier density (\nloc) as determined from resistivity~\cite{Pelc19,Tabis21,NBarisic22}. Its variation from 1 in PG-FL regime to 0 in OD-FL is represented by the color scale. The solid black and dashed grey isolines track its evolution. The superconducting (SC), and antiferromagnetic (AF) regions are also shown.  \Ts\ corresponds to $\nloc \sim 97\%$, whereas \Tss corresponds to $\nloc\sim99\%$. The blue (red) chevrons indicate the doping of the Hg1201 (Bi2212) samples discussed in the text. (e) Comparison of the zero-temperature effective carrier density  \neff\ for several cuprates families extracted from the resistivity (\nrho, full lines) and Hall effect (\nH, filled circles)~\cite{Pelc19,Tabis21,NBarisic22,Putzke21}. Considering the uncertainties in the absolute values of $\rho$ and \RH, and the difficulties associated with ascertaining the exact doping level for each particular sample, the agreement between \nrho\ (extracted first by use of a simple model \cite{Pelc19}) and \nH\ (directly measured by use of high magnetic fields somewhat later \cite{Putzke21}) across different compounds is remarkable. 
    }
    \label{Fig1_dc_resistivity_Phase-diagram}
\end{figure*}

The discovery of high-temperature superconductivity in cuprates, a phenomenon that no one except their discoverers expected in these compounds~\cite{Bednorz86}, is among the most important breakthroughs in solid-state physics. This family of materials has been a continuous source of surprises with unconventional, sometimes compound-specific, behaviors inspiring numerous and exciting, often mutually exclusive, ideas and theories~\cite{Keimer15}. At half-filling, the cuprates are charge-transfer insulators~\cite{Zaanen85}, with one localized hole per CuO$_2$ unit (Fig.~\ref{Fig1_dc_resistivity_Phase-diagram}). At $p\approx 0.25$ doping, corresponding to 0.25 added holes per CuO$_2$ unit, the so-called overdoped regime, they are a conventional Fermi liquid (FL) with an effective carrier density $\neff=1+p$. Superconductivity (SC) appears between $p\approx 0.05$ and $\approx 0.25$ doping, with a maximum \Tc\ at optimal doping around $p\approx 0.16$. 

This general pattern is observed for many structurally distinct cuprates, which indicates a universal physical origin of SC. In particular, it has been demonstrated that the normal-state sheet resistance is essentially universal~\cite{NBarisic13, NBarisic19}. Optimal doping is characterized by a linear-like temperature dependence of the resistivity, which persists from very high to surprisingly low temperatures~\cite{Gurvitch87}. Along with the unusual evolution of spectroscopic responses, it was generally interpreted as indicative of exotic, non-FL behavior~\cite{Marel03,Keimer15}. In the underdoped regime, this linear-like dependence is observed only at high temperatures. The low-temperature deviation from linearity appears below a temperature \Ts\ that delimits the pseudogap regime from above. It is precisely in the pseudogap regime that the entirely unexpected FL nature of the itinerant charge was clearly revealed by its unambiguous theoretical signatures. The resistivity is quadratic in temperature ($\rho=A_2 T^2$) and inversely proportional to doping ($A_2\propto 1/p$)~\cite{NBarisic13,Ando04a}; the Hall coefficient is temperature-independent and identifies the carrier density to be $p$ ($\RH\propto 1/p$)~\cite{Ando04,NBarisic19}; the inverse of the Hall mobility is consequently also quadratic in temperature ($1/\muH= C_2T^2$); the magnetoresistivity scales as a function of field and temperature according to Kohler's rule, which reveals that the scattering rate is of FL nature as well~\cite{Chan14}. Finally, the frequency and temperature dependence of the optical scattering rate also scale in the manner expected from a FL~\cite{Mirzaei13}, with a small but significant modification in the scaling variable, which is fully explained in this article. 

%
%
Because the pseudogap regime is central to the cuprate problem, the discovery that the itinerant charges in it are a true FL provides a solid fulcrum around which the whole interpretation of cuprates can be understood. The lever of the fulcrum in this analogy is the complete universality of the Hall mobility, a key experimental fact overlooked until quite recently~\cite{NBarisic15,NBarisic19}. In contrast to resistivity, which changes from quadratic in temperature to linear-like at optimal doping, and then back again to quadratic in the overdoped regime, the Hall mobility remains virtually unchanged throughout the temperature-doping phase diagram. Furthermore, it is essentially compound-independent, quadratic in temperature ($1/\muH= C_2 T^2$) with the same coefficient [$C_2=0.0175(20)\,{\rm T}{\rm K}^{-2}$] across all cuprates~\cite{NBarisic15,NBarisic19}. The implication, in agreement with the observed evolution of the arcs at the Fermi surface, is that the ungapped (nodal) parts retain their FL character with an essentially universal scattering rate and Fermi velocity (effective mass), while the gapped (antinodal) states do not contribute to planar conductivity~\cite{NBarisic15,NBarisic19}. Consequently, the deviations from the FL quadratic temperature dependence in the resistivity can be simply related to the change in the carrier density, obtained from the resistivity directly by use of a standard Drude ($\tau(T,\omega=0$)) formula, and associated with the repopulation of the gapped parts of the FS. In this way, one can count the itinerant charges throughout the phase diagram~\cite{Pelc19}. We denote the mobile charge concentration, extracted from the measurements, with \neff. Its evolution at $T=0$ is shown in Fig.~\ref{Fig1_dc_resistivity_Phase-diagram}e. 

Once \neff\ is unambiguously identified from the resistivity~\cite{Pelc19} or Hall coefficient~\cite{Putzke21}, keeping in mind that the total available charge is always $1+p$, one finds the evolution of the localized charge concentration \nloc\ across the whole phase diagram from simple charge balance~\cite{Pelc19}: 
\begin{equation}
    1+p=\neff + \nloc.
    \label{eqn:neffnloc}
\end{equation}
Because \nloc\ is fixed by the other two known quantities, it is unambiguous as well. It is presented as a function of temperature and doping in Fig.~\ref{Fig1_dc_resistivity_Phase-diagram}d. Within the pseudogapped FL (PG-FL) regime in particular, exactly one hole  is localized per unit cell ($\nloc=1$) below the temperature \Ts\ and is associated with the missing part of the Fermi surface (FS). The remaining $\neff=p$ holes, determined precisely by transport measurements~\cite{NBarisic15,Ando04,Padilla05}, are therefore associated with well-formed arcs. 

In principle, optical conductivity measures both itinerant and localized charges directly. Nevertheless, no consensus has been reached so far about the meaning of the optical conductivity data. In particular, at optimal doping the frequency dependence of the optical response is featureless up to about 1 eV~\cite{Bozovic90}, which was therefore fitted and interpreted in numerous ways, thus becoming one of the cornerstones of ill-defined non-FL approaches~\cite{Collins89,Puchkov96,Anderson97,Marel03,Heumen09}. 

Here, we resolve this decades-long conundrum. Understanding that the transport measurements necessarily account for the low-frequency part of the optical conductivity, we have essentially no freedom in fitting the optical conductivity, in contrast to previous two-component analyses~\cite{Collins89,Kamaras90,Liu06,Padilla05}. Conversely, optics provides independent information on that part of the response which is not directly involved in transport. We implement this idea by noting that the FL part of the optical response can always be unambiguously reconstructed from FL transport data, because a FL has theoretically well-defined responses for all observables. Consequently, the FL optical response is entirely determined by the effective mass \ms, carrier density \neff, and relaxation time $\tau$ as measured by transport at $\omega=0$. We then compare the  calculated FL optical response with the actual optical measurements. Observing that the low-energy part of the optical conductivity is reconstructed correctly, we demonstrate that the optical responses of cuprates are fully consistent with FL transport. The remainder identifies the experimental spectral responses related to the localized charge essentially without any fitting or ambiguity. 

We perform this analysis in the PG-FL regime first, where the localized-itinerant duality of the cuprates reveals itself most clearly, as implied by pure $\rho \propto T^2$, e.g., in Fig.~\ref{Fig1_dc_resistivity_Phase-diagram}b. For this purpose, we choose the single-layer compound HgBa$_2$CuO$_{4+\delta }$ (Hg1201), where the FL properties were clearly established by previous work~\cite{NBarisic13,Mirzaei13,Chan14,NBarisic19}. Second, by analyzing the corresponding data for the structurally more complex double-layer Bi$_2$Sr$_2$CaCu$_2$O$_{8+\delta }$ (Bi2212), we demonstrate that this approach naturally carries over to optimal doping, where $\rho \propto T$, see Fig.~\ref{Fig1_dc_resistivity_Phase-diagram}c. Third, all the insights gained in the PG part of the phase diagram apply to the overdoped part as well (in agreement with the universality of \muH), based on direct observation of evolution in the transport \neff, from $\neff=p$ to $\neff=1+p$. Fourth, we note that the loss of the FL spectral weight associated with superconductivity in the underdoped regime tracks \neff~\cite{Homes04,Homes05,Pelc19}, while in the overdoped regime it tracks the loss of the localized charge \nloc~\cite{Mahmood19,Pelc19,NBarisic22}. Finally, we show that the same method of separating the optical response into FL and non-FL contributions may also be applied to superconducting pnictides and to Sr$_2$RuO$_4$, although the physical origin of the non-FL part is different than in the cuprates: it is caused by dissipative electronic pockets with a low energy scale $\delta$. For Sr$_2$RuO$_4$, $\delta \approx 5$~meV, whereas for the pnictides we conclude that it is even smaller. Our work thus demystifies the seemingly complicated optical response in cuprates, as well as in a wide class of related correlated systems.


\section{Results}
\label{sec:results}

%
%
\begin{figure*}[!htb]
    \centering
    \includegraphics[scale=0.46]{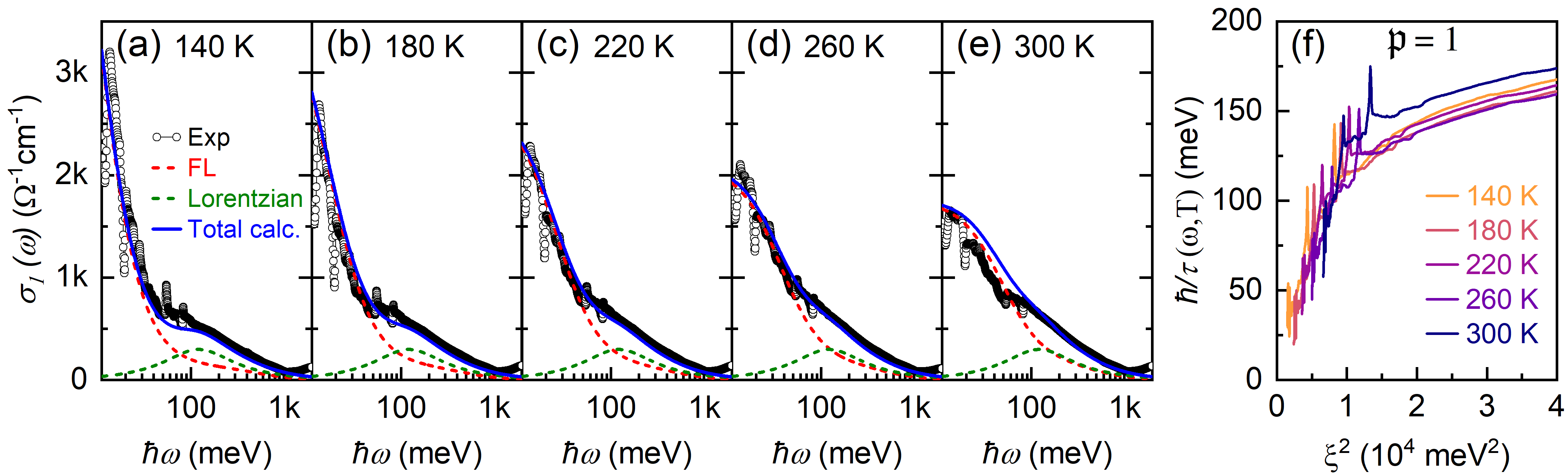}
    \caption{(a--e) FL contribution to optical conductivity of underdoped Hg1201 ($p\approx 0.1$) calculated from transport using Eq.~(\ref{AP01gr}) (red dashed curve) and a temperature independent Lorentzian peak centered at 125~meV (green dashed curve) captures the measured optical conductivity (circles) well for all temperatures (the blue curve).  (f) Good scaling of the overall optical scattering rate $1/\tau (\omega )$ [Eq.~\eqref{eqn:Domega}], determined from the experimental data presented in (a--e), is obtained for $\pp=1$. Deviation from $\pp=2$, expected for a FL, is simply due to the presence of two contributions (FL with $\pp=2$ and MIR-Lorentzian) that are mixed by the extended Drude analysis. The scaling works down to lowest energies, as expected when the ground state is interpreted correctly. Here $\ms=3.5\,m_e$, $C_2=0.016\, ~{\rm T}{\rm K}^{-2}$~\cite{NBarisic19}. }
    \label{Fig:Hg1201OptCond}
\end{figure*}


\subsection{Reconstruction of a Fermi liquid optical response}
\label{sec:FL_reconstruction}

A characteristic signature of the Fermi liquid (FL) is the scaling of the frequency- and temperature-dependent optical scattering rate $\tau (\omega,T)$. In the limit of well-defined quasiparticles, the response is a function of the single scaling variable~\cite{Berthod13}
\begin{equation}
    \xi^2=(\hbar \omega )^2+(\pp \pi \kB T)^2,
    \label{eqn:xisqr}
\end{equation}
where the scaling parameter $\pp=2$ exactly for the FL case. In practice, it is well documented that the scaling parameter $\pp=2$ is not observed experimentally even in elemental metals. In several correlated metals, when \pp\ is treated as an effective parameter, its best-fit values range from $1$ to $2.4$~\cite{Sulewski88, Katsufuji99, Yang06, Dressel11, Nagel12, Mirzaei13,Stricker14,Reber19,Wang21,Pustogow21}. The observed scaling collapse is usually quite good and the departure from the FL value is typically attributed to unusual scattering mechanisms and non-FL behaviors. To compare with experiments, and assess the importance of non-FL corrections to the optical response of cuprates, we first determine the effective $\pp$ by applying an extended ($\tau(T,\omega\geq 0$)) Drude analysis, as if the optical response were all due to some itinerant charges.

Next, we introduce an alternative to the approaches based on the effective parameter \pp. For cuprates, we consider a straightforward interpretation of the experimental data under the assumption that itinerant and localized charges give two different contributions. Regarding the former, we use previously derived theoretical expressions for the optical response of the FL~\cite{Berthod13}. In particular, we calculate the expected optical response using \neff\ and $\tau$ obtained from transport measurements as input (see Appendix~\ref{AppA}), with no fitting parameter ($\pp=2$). In that case, there appears, in addition to this FL contribution, an additional MIR feature associated with the localized charge. In this way, the departure of \pp\ from its FL value is fully explained without ascribing non-FL behavior to the itinerant charges: of course, the localized charges are not a FL, so they cause deviations from $\pp=2$ when subsumed in the extended Drude analysis for $\tau(T,\omega)$. As argued below, an extension of this idea is also applicable to pnictides and strontium ruthenate, and probably to other functional ionic materials as well. 


\subsection{Optical response in the underdoped regime}
\label{sec:OptReponse_UDregime}

The PG regime is a good starting point for the analysis of conductivity, as argued in the Introduction. Hg1201 is a model system for experimental studies because of its simple tetragonal crystal structure, minimal point disorder effects, and record optimal \Tc\ among single-layer cuprates~\cite{Eisaki04,NBarisic08,Chan14}. As shown in Fig.~\ref{Fig1_dc_resistivity_Phase-diagram}, the resistivity of the measured sample in the PG regime is purely quadratic, as established long ago~\cite{NBarisic08,NBarisic13}. Figures~\ref{Fig:Hg1201OptCond}(a)-\ref{Fig:Hg1201OptCond}(e) shows the frequency-temperature dependence of $\sigma_1$ (see the Appendix~B). We emphasize that the dc and optical conductivity data were obtained on the same sample to minimize systematic errors in the data analysis. 

The theoretical FL frequency-temperature dependence of $\sigma_1$ is calculated from transport (blue dashed line in Fig.~\ref{Fig:Hg1201OptCond}, see also Appendix~B), while the remainder of the spectrum is fitted to a temperature-independent Lorentzian after subtracting the calculated FL part. Even at the highest-measured temperature, there is no need for any corrections because the increase in carrier density, \neff [Fig.~\ref{Fig1_dc_resistivity_Phase-diagram}d] is only of the order of a percent. Nevertheless, one can observe an indication of the delocalization process as a small overestimation of the overall fit with respect to the data collected at 300~K [Fig.~\ref{Fig:Hg1201OptCond}e].  This small correction is hardly  captured by optics, but it can be detected clearly by resistivity, as a change from quadratic to linear-like temperature dependence [Figs. \ref{Fig1_dc_resistivity_Phase-diagram}(a) and \ref{Fig1_dc_resistivity_Phase-diagram}(b)]. By applying the extended Drude analysis, the ``overall optical scattering rate'' (i.e., a mixture of the contributions from both subsystems) is calculated for each temperature from the measured $\sigma_1$ and by using a plasma frequency, $\omega_p$, that corresponds to \neff\ at 200 K. It is shown to scale with $\pp=1$, as in Fig.~\ref{Fig:Hg1201OptCond}f and the Appendix~B. Of course, the \pp\ for the FL part is exactly 2 (see Appendix~B). Clearly, unlike the FL scattering rate, the overall scattering rate is an \textit{effective} quantity without a clear physical meaning, as it incorporates both the FL and localized (``non-FL'') contributions. 

%
%
\begin{figure*}[!htb]
    \centering
    \includegraphics[scale=0.52]{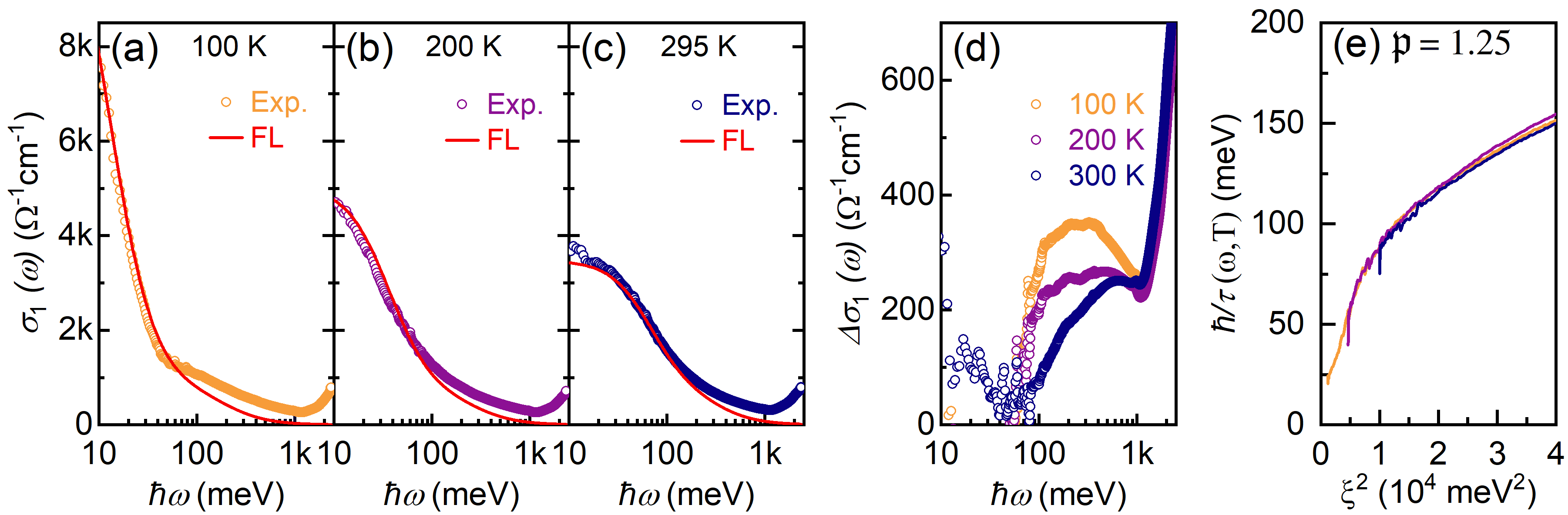}
    \caption{(a)-(c) Experimental optical conductivity (open circles)~\cite{Tu02} and calculated FL contribution (red line) of optimally doped Bi2212 ($p\approx0.16$) at selected temperatures ($T>T_c$). The calculated FL contribution clearly captures well the low-frequency part of the conductivity spectra for all temperatures. (d) The MIR non-FL contribution to the optical conductivity is obtained by subtracting the FL contribution from the total optical conductivity. Its temperature evolution clearly reveals the spectral weight transfer from MIR to FL with increasing temperature. This transfer is consistent with thermal excitation of localized carriers (\nloc) and inconsistent with dissipation scenarios. (e) Because of the MIR contribution, the overall scattering rate scales with $\pp\approx1.25$. Here $\ms=3\,m_e$, $C_2=0.021\,~{\rm T}{\rm K}^{-2}$~\cite{NBarisic19}. 
    }
    \label{Fig:Bi2212OptCond}
\end{figure*}

The doping dependence of the optical response in underdoped La$_{2-x}$Sr$_x$CuO$_4$ (LSCO) and YBa$_2$Cu$_3$O$_{6+x}$ (YBCO) was successfully fitted at a single temperature by a sum of standard Drude and higher-frequency Lorentz contributions already some time ago~\cite{Padilla05}. That analysis also showed that \ms\ was compound- and doping-independent, and that $\neff=p$. However, such a two-component fitting at a single temperature could not be extended beyond the PG regime, because \neff\ and \nloc\ start to overlap strongly with increasing doping, and finally, in the overdoped regime, the optical response is dominated by \neff. The analysis conducted here is qualitatively and quantitatively fully constrained and can be applied at all dopings, because the FL component is fixed by an independent dc transport measurement, which reveals the MIR feature unambiguously, even at overdoping where it is small and appears at low frequency. Because the FL turns out to account for the low-frequency part of the optical response at all temperatures, it is clearly universal.

%
%
\subsection{Optical response at optimal doping}
\label{subsec:OptDop_OptResp}
Optimal doping is a delicate crossover regime. The reason is clear: according to the analysis of transport properties~\cite{Pelc19}, at this doping \neff\ and thus \nloc\ changes most rapidly with temperature and doping [Fig.~\ref{Fig1_dc_resistivity_Phase-diagram}d]. This crossover accounts for the linear-like temperature dependence~\cite{Pelc19}. In this regime, we choose to analyze Bi2212. This well-studied cuprate features considerably lower structural symmetry than Hg1201, and two rather than one CuO$_2$ planes per formula unit. Its resistivity is linear essentially down to \Tc, as shown in Fig.~\ref{Fig1_dc_resistivity_Phase-diagram}c. We analyze only previously published data~\cite{Tu02, Kendziora92}, testifying to the directness and consistency of our approach. 

The analysis is shown in Fig.~\ref{Fig:Bi2212OptCond}. Just like for underdoped Hg1201, the expected FL contribution is calculated from the dc transport data in Fig.~\ref{Fig1_dc_resistivity_Phase-diagram}c. The low-frequency part of the spectrum is perfectly matched at all temperatures [Figs.~\ref{Fig:Bi2212OptCond}(a)-\ref{Fig:Bi2212OptCond}(c)]. The extracted MIR weight decreases with increasing temperature, consistently with a transfer of charge from the localized to the itinerant subsystem, corresponding now to a $\sim$10--20\% increase of $\neff$ above $p$, in contrast to $\sim$1--2\% in Hg1201 [Fig.~\ref{Fig1_dc_resistivity_Phase-diagram}d]. As in the case of underdoped Hg1201, approximate scaling can be obtained for the overall optical scattering rate. For Bi2212, the effective value that yields the best scaling is $\pp\approx 1.25$ [Fig.~\ref{Fig:Bi2212OptCond}e and Appendix~B]. The FL part alone scales with $\pp=2$ as always. 


\subsection{Optical response in the overdoped regime}
\label{subsec:OvrDop_OptResp}
In the overdoped regime, the FL contribution is much larger than that of the localized charge, simply because the delocalization is almost complete~\cite{Pelc19}, and thus the MIR spectral weight is small. This makes the direct observation of the localized response difficult. However, much can be concluded just by following the doping evolution of the MIR feature. As repeatedly pointed out earlier~\cite{Pelc19,NBarisic22}, it is obvious just by looking at the data (Ref.~\cite{Uchida91} and Appendix~B) that the spectral weight of exactly one charge per CuO$_2$ unit is transferred to the FL response from the higher energies. The MIR peak first increases the more it approaches the FL peak, to be absorbed by it in a unitary FL response of weight $1+p$ only at extreme overdoping. 

\begin{figure}[!t]
    \centering
    \includegraphics[scale=0.34]{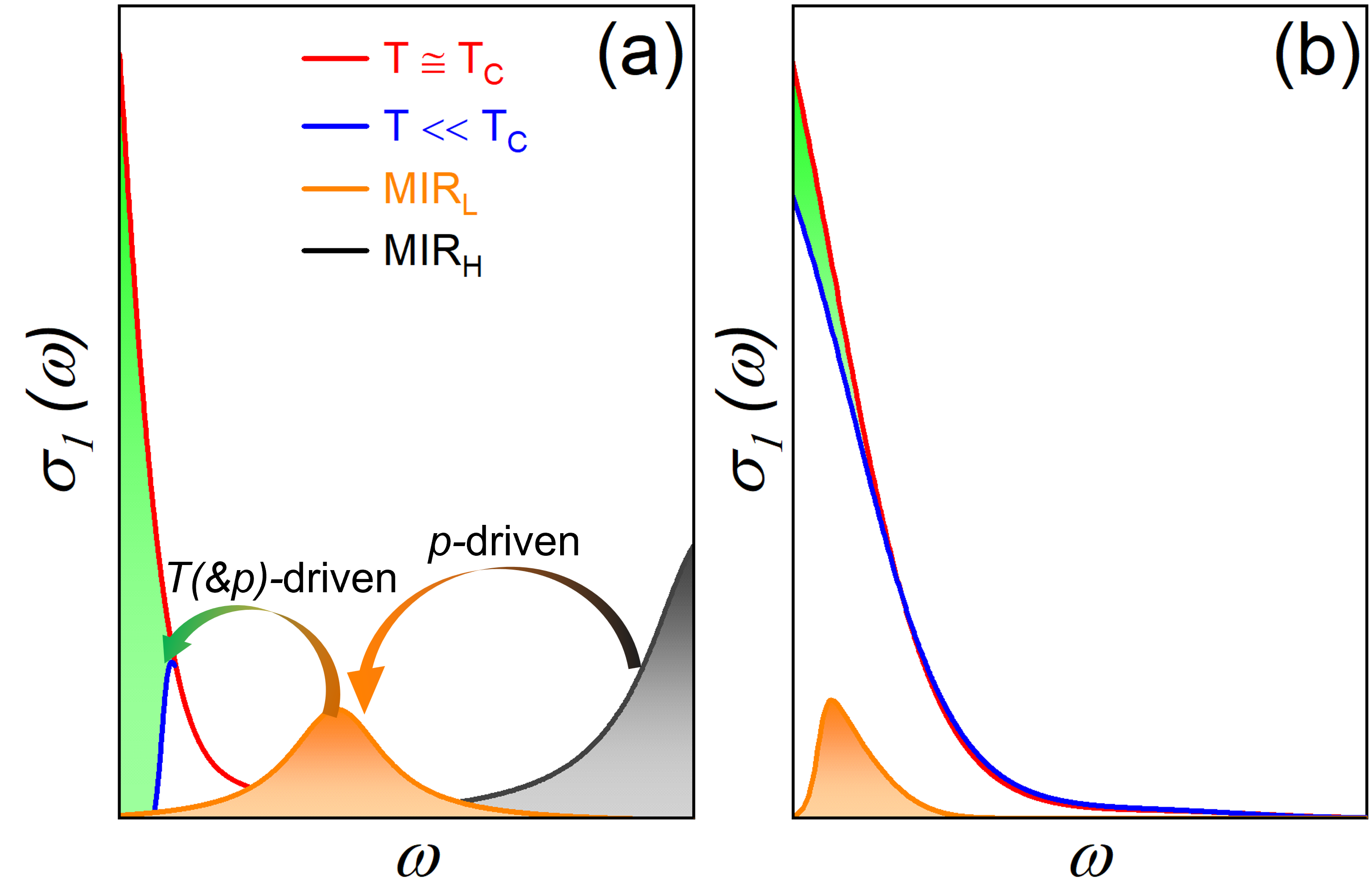}
    \caption{A schematic representation of the temperature and doping evolution of the optical spectral weight in cuprates. (a) In the normal state, for $T>\Tc$, the spectrum consists of three contributions: FL part (indicated with a red line) that corresponds to \neff, MIR feature (orange) which is associated with a part of \nloc, and spectral weight located at high energies (above charge transfer gap), associated with the remaining part of \nloc. Upon doping, the MIR feature shifts to lower frequencies, growing in weight at first, to be absorbed by the FL peak in the highly overdoped regime. It brings in a spectral weight of exactly one charge per CuO$_2$ unit. The dominant effect of increasing the temperature is the spectral weight transfer from the MIR to the FL. Upon decreasing the temperature below $\Tc$, the spectral weight, corresponding to the green area, is shifted from the FL part of the spectrum to the SC delta-function peak at $\omega=0$, yielding the Homes law~\cite{Homes04,Homes05}. (b) In the overdoped regime, the FL part of the spectrum is only partially SC-condensed (green area), while the non-condensed spectral weight corresponds to insufficient ``glue'' \nloc~\cite{Pelc19}.
    }
    \label{Fig:SpectWtSchem}
\end{figure}

In contrast to optics, transport is a delicate probe, with $\kB T$ resolution at zero frequency, so that it reveals precisely where (de)localization begins, along with its full evolution with doping. Both the resistivity and Hall coefficient identify the same doping dependence of \neff\ and thus \nloc\ [Figs.~\ref{Fig1_dc_resistivity_Phase-diagram}(d) and \ref{Fig1_dc_resistivity_Phase-diagram}(e)]. Another important observation is that the superfluid density \rhoS, which corresponds to the missing spectral weight in the FL contribution below \Tc\ (see Appendix~B), becomes limited by \nloc~\cite{Pelc19}. This behavior implies that \nloc\ is associated with an excitonic-like (short-range, short-time) glue~\cite{NBarisic22}. Consequently, 
\begin{equation}
    \rhoS=\left( O_{\rm S}~\nloc \right) \neff,
    \label{eqn:rhoS}
\end{equation}
a relation which notably holds across the superconducting dome~\cite{Pelc19}, with a material-dependent constant $O_{\rm S}$ discussed elsewhere~\cite{NBarisic22}. It should be noted that \rhoS\ is proportional to the mobile concentration $\neff\propto\omega_p^2$ here, with the constant of proportionality containing \nloc\ as the probability to encounter the localized hole, which provides the glue. Therefore, some normalizing denominator is expected in $O_S$, such as $O_S\propto 1/(\neff + \nloc)$, or simply a constant (per unit cell). All the above experimental facts and reasoning are schematically summarized in Fig.~\ref{Fig:SpectWtSchem}.

%
%


\section{Discussion}
\label{subsec:DiscussConclude}

The experimental results described in this work are the keystone of a simple scenario for both the normal and superconducting states in the cuprates, which rests on an unambiguous experimental determination of the actual concentration of mobile (FL) charges \neff\ present in a sample. The implication of a second, localized charge subsystem which balances Eq.~\eqref{eqn:neffnloc} for the total number of charges $1+p$ begs the question: Which spectroscopic signatures characterize the localized subsystem? The most prominent of these is the mid-infrared (MIR) peak, studied in this work at optimal doping and in the PG regime. 

The MIR peak is most pronounced in the PG regime because the two electronic subsystems \neff\ and \nloc\ are still reasonably well separated in frequency there. In Fig.~\ref{Fig:Hg1201OptCond}, the presence of the MIR peak explains why the Fermi-liquid scaling parameter $\pp$ for the total (measured) value of the optical response is not equal to its textbook value $\pp=2$. Quite simply, it is because the localized charges do not conduct, so it is not physical to treat the total response as a FL response~\cite{Allen77}, which is exactly what the extended Drude analysis presumes. Taking into account the FL contribution alone, one recovers $\pp=2$. Because the FL peak is independently obtained from transport, essentially without any fitting parameters, its subtraction from the optical response defines the MIR response experimentally, which can then be fitted with a Lorentzian (or any similar functional shape). In this way, the MIR feature is fully characterized, and tied to the localized sector of the charge-balance equation~(\ref{eqn:neffnloc}). 

Increasing the temperature (or doping) across the \Ts\ line leads to a gradual delocalization of the hole. The \Ts\ (PG) line in Fig.~\ref{Fig1_dc_resistivity_Phase-diagram}(e) corresponds to the line of constant $\nloc\approx 0.97$~\cite{Pelc19}. For Hg1201, this beginning of delocalization at the percent level falls below the experimental resolution of the optical conductivity, and thus can not be easily resolved in the evolution of both FL and MIR contributions (Fig.~\ref{Fig:Hg1201OptCond}). This observation solves yet another long-standing mystery in cuprates, namely, why the \Ts\ line is not observed in the in-plane measurements of optical conductivity. The dominant effect is the simple temperature broadening of the FL peak. 

The optical response of Bi2212 at optimal doping, Fig.~\ref{Fig:Bi2212OptCond}, considered by many as the most mysterious regime, repeats the same pattern. In contrast to Hg1201, which is paradigmatic among cuprates for its simplicity, Bi2212 is among those exhibiting more complex behavior. Thus their parallel behavior is significant experimental support of the presumed universality. Because the localized hole delocalizes more quickly with the temperature here [Fig.~\ref{Fig1_dc_resistivity_Phase-diagram}d], one can observe the concomitant spectral-weight transfer with the naked eye [Fig.~\ref{Fig:Bi2212OptCond}d]. Notably, the FL spectral weight as deduced from resistivity---the area under the coherent peak---increases while the MIR weight decreases with {\it{increasing}} the temperature, as clearly observed in Fig.~\ref{Fig:Bi2212OptCond}d. Such a spectral-weight transfer is expected in the case of temperature excitations across a gap (e.g., in a semiconductor), in contrast to what is expected if part of the itinerant \neff\ should become incoherent with temperature. Namely, if the MIR peak reflected an incoherent (dissipative) structure, it would be expected to increase with the temperature, not decrease as observed. Here again, the overall (effective) Fermi liquid scaling parameter $\pp\approx 1.25$ is a simple consequence of the MIR contribution in the extended Drude analysis. 

A gapped spectral feature cannot give rise to dissipation at temperatures below the gap (see Appendix~B). In this way, our measurements clearly eliminate a dissipative mechanism for the disappearance of a part of the Fermi surface in the PG regime. There is similarly no indication of a quantum critical regime anywhere in the phase diagram: \ms\ (or $v_F$) \cite{Padilla05,Zhou03} and $\tau$ \cite{NBarisic15,NBarisic19} remain unchanged, while the evolution of \neff\ is smooth throughout the overdoped region\cite{NBarisic19,Pelc19}. Notably, the  universality of the absolute values of the overall $\tau$, displayed in Figs.~\ref{Fig:Hg1201OptCond}f and ~\ref{Fig:Bi2212OptCond}e, obtained directly from $\sigma (\omega )$ by using \neff, confirms those experimental observations independently. The smooth evolution of \neff\ with doping implies that the complementary density \nloc\ behaves like the aggregate variable of a first-order or percolative (power-law) phase transition, with no indication of the non-vanishing long-range correlations which accompany second-order (quantum) phase transitions~\cite{NBarisic22}.

In our view, the non-Fermi or marginal Fermi liquid concepts simply reflect confusion about the number of itinerant charges \neff\ present at various doping levels. Notably, one can account for the Fermi arcs even within a strictly one-body DFT+\emph{U} model~\cite{Lazic15}. Once the 3D background of dopant Coulomb fields is properly included, they produce a FL with nodal arcs growing with $p$, as observed, while the antinodal parts of the FS remain in the original (gapped) charge-transfer parent state. Hence the mere observation of the arcs and PG does not in any way require going beyond the ordinary FL picture for the itinerant charges. Conversely, including static charges in the transport accounting would make any metal look like a ``bad metal.''

As shown in detail in Appendix~B, these considerations can be brought to bear fully and fruitfully to the other two notable ``bad metal'' groups, pnictides and ruthenates, and are also relevant for other families of materials ~\cite{Sulewski88, Katsufuji99, Yang06, Dressel11, Nagel12}. The salient point is that the parameter \pp\ in the FL scaling variable in Eq.~\eqref{eqn:xisqr} is {\it{effective}}, meaning simply that using it as a free parameter gives very good scaling fits from zero to the MIR range. In general many-body theory, the initial approach (lowest diagram) gives the structure of a formula, while higher-order corrections give formulas for the parameters in the initial expression. Hence the effectiveness of \pp\ in the above-mentioned examples of ``bad metals'' proves that the FL is the correct starting point for them too: they are not really bad. The reason for the departure of the effective  \pp\ from $\pp = 2$ varies from case to case, and this can be used to reveal the physical mechanism behind the corrections by modeling.

In the cuprates, as already noted, the correction comes from the MIR feature, which reflects the non-conducting localized hole. In the pnictides (see Appendix~B), aside from the FL contribution, there is a constant contribution down to lowest frequencies~\cite{NBarisic10}, as if from a subsystem of fixed scatterers. It does not necessarily follow that they contain impurities. The constant contribution is more probably due to a flat band with a maximum (X-point van Hove singularity) pinned at the Fermi level~\cite{Derondeau17}, which is incoherent because it is smeared out even by a small temperature. Similarly, in ruthenates, a low-energy electronic pocket in the spectrum, with a scale $\delta \approx 5$~meV, was clearly observed by photoemission (see Appendix~B). It is to be expected that, as soon as $\kB T>\delta$, the pocket becomes an extraneous source of dissipation to the remaining FL. And indeed the resistivity up to ${\approx}40$~K (${\approx}5$~meV) is quadratic in temperature, while the scattering rate scales with $\pp=2$. Crossing this temperature, resistivity becomes sub-quadratic (non-FL, $\rho=AT^n$, where $n<2$), to become linear-like at high temperatures. Concomitantly, \pp\ begins to decrease towards 1. In our interpretation, dissipation clearly behaves as expected, vanishing at low temperatures. To make the same point conversely, if the PG part of the FS in cuprates were due to a dissipative mechanism, the Fermi arcs would grow and touch the edge of the zone as $T\to 0$. 

The only remaining issue is that the normalized integral under the MIR feature is not unity. For example, at 140~K in Hg1201 with $p\approx0.1$ it corresponds to ${\approx}0.12$ holes per CuO$_2$ unit. Thus, part of the spectral weight related to \nloc\ remains at higher frequencies, probably where it was in the parent compound. There are two complementary ways to understand that, based on describing the localized hole as a locally ionic polaron-like structure, named the $d$~complex and extensively discussed in Ref.~\cite{NBarisic22}. It consists of a planar CuO$_4$ ``molecule'' and its apical oxygens where applicable, along the lines originally suggested by M\"uller~\cite{Mueller05}, who emphasized that the symmetry of this local deformation could correspond to any of the known macroscopic phases (HTT/LTT/LTO) associated with oxygen octahedra. Such structural disorder is documented extensively in cuprates. It is therefore natural to assume that the observed MIR peak corresponds to only one of these deformations, while the others could easily hide underneath the large rise in optical conductivity as one approaches the higher-energy (IR) region, the beginning of which is clearly visible in Fig.~\ref{Fig:Bi2212OptCond} (see also Appendix~B). Second, a spectroscopic probe in principle excites the whole spectrum of a structure, so if the MIR peak is the first excited state of the $d$~complex, the higher excited states should similarly be draining spectral weight, hidden under the high-energy IR plateau. To clarify this important issue, a similar analysis of optical conductivity should be conducted at various dopings in different compounds. In direct support of these scenarios, the weight of the MIR feature grows as it approaches zero energy at overdoping, clearly dominating excitations at all higher energies [Fig.~S11 in Appendix~B].

The structure of Eq.~\eqref{eqn:rhoS} encapsulates the whole SC dome~\cite{Pelc19}, because \neff\ rises with $p$ at (nearly) constant \nloc\ until optimal doping. At overdoping \neff\ rises quite rapidly from $p$ to $1+p$, but \nloc\ drops to zero in compensation, according to Eq.~\eqref{eqn:neffnloc}. Because \neff\ is the density of mobile charges, $O_{\rm S}\nloc$ plays the role of coupling strength or glue. The disappearance of SC on the overdoped side is then simply interpreted as loss of the glue provided by localized holes~\cite{Pelc19,NBarisic22}. Thus the latter explains why the fraction of the spectral weight of the FL condensed in \rhoS\ decreases even as the density of mobile charges \neff\ increases: \rhoS\ is proportional to the vanishing glue \nloc.

It should be emphasized that our work not only demystifies all the well-known aspects of cuprate physics, debated for decades, as discussed above, but also provides rather straightforward explanations for several most recent experimental observations, some of which appeared after the initial submission of this paper. We note with pleasure that more and more groups across the globe are beginning to realize that the responses they are measuring imply the presence of a Fermi-liquid electronic (sub)system. For example, it was deduced from studies of the high-field in-plane magnetoresistance~\cite{Ayres21, Grissonnanche21} that one component is “Planckian” while the other is classical, or it was argued that all the salient cuprate's feature can be discussed as a system with localized “strange scatterers" embedded into a Fermi liquid with a conventional quasiparticle description~\cite{Berg22}. Nevertheless, confusion in the field remains~\cite{Phillips22}, caused by conflating the effects of the \emph{scattering rate} with changes in the \emph{carrier density}, as well as by a lack of distinction between \emph{coherency}, \emph{incoherency}, and \emph{localization}, all of which is completely clarified by our work here.

Because of this confusion, even when clear and unambiguous signatures of standard Fermi-liquid behavior are observed, the itinerancy is still discussed in the context of non-Fermi liquid physics. An excellent example (out of many) is the recent finding that electrons at optimal doping obey standard (FL) orbital motion in a magnetic field~\cite{Ataei22}. Nevertheless, this was attributed to Planckian physics, though an entirely natural explanation is to associate them with the FL subsystem, as revealed by optics and transport in the present work.

The same confusion persists in the analysis of optical conductivity data. A good example is the very recent analysis of optical conductivity in BSCO (Bi$_{2-x}$Pb$_x$Sr$_{2-y}$La$_y$CuO$_{6+\delta}$), where the coherent contribution is simply fitted by a Drude ($\tau(T,\omega=0)$) term, overlooking the observed transport universalities~\cite{Heumen22}. Even within such a crude approach, it was reported that the Drude weight is quite weakly temperature-dependent and changes by at most 5\% up to room temperature for overdoped samples, while with decreasing doping the temperature dependence increases somewhat, but never exceeds 15\%, which agrees well (despite claims to the contrary) with what was deduced based on transport [Fig. 1] and what is documented in Fig. 3. At optimal doping the increase of the Fermi-liquid spectral weight, from \Tc\ to room temperature, corresponds to a $\sim$10--20\% increase of \neff. As argued here, this increase of spectral weight with temperature is a certain signature of the (second) \emph{gapped} electronic subsystem (MIR). However, the Drude weight in BSCO evolves monotonically throughout the entire doping range studied, and the universal $p$ to $1 + p$ crossover in carrier density, seems to be absent in this compound. In the present context, such an observation is not very surprising, because it is well known that in BSCO, as well as in LSCO~\cite{Ino02}, a flat band or extended van-Hove  singularity is pinned to the Fermi level in the crossover doping range between $\neff=p$ and $1+p$~\cite{Takeuchi01,Kondo04}. Saddle points very close to the Fermi level will result in an \emph{incoherent} response, as discussed here in the case of iron-pnictides, and thus will not be captured by a \emph{coherent} Drude term. Nevertheless, the  scattering rate extracted by Drude fittings to the residual term shows a roughly doping-independent increase with temperature. This result is in agreement with the universality of transport and the determined optical scattering rate. It should be also noted that the observation of Mathiessen's rule is an unambiguous signature of quasiparticles, and there is no reason for it to mean anything else in cuprates. The probability of quasiparticle scattering adds in the resistive (serial) channel, while contributions from both incoherent and (pseudo)gapped parts of the Fermi surface add in the conductive (parallel) channel, hence the FL easily overwhelms the non-FL contribution in transport.

Notably, if one is not interested to understand the scaling parameter \pp, even a conventional Drude term [with $\tau(T,\omega=0$) and $\ms=$~\textit{const.}] is sufficient to capture the temperature and frequency dependence of the optical conductivity.  The difference between Drude, extended-Drude [$\tau(T,\omega$) with effective $\pp$], and Fermi liquid (with $\pp=2$) terms is important only for the discussion of the frequency dependence of the optical scattering rate. Thus, in order not to lose an important part of the cuprate physics, the Fermi-liquid term was used here.

The phase diagram is essentially universal for all cuprates. Relying on similarly universal electronic properties, we have clarified it broadly in full. However, each cuprate compound also exhibits different (maximal) superconducting transition temperatures, roughly categorizing them into high and low-T$_c$ compounds. This difference must stem from a non-universal component of the cuprate problem. It was demonstrated that the oxygen degrees of freedom, tuned by the non-universal local-tilt symmetry of the lattice (HTT/LTT/LTO), affect the superconducting properties directly~\cite{Barisic90,NBarisic22,Zeljkovic12,Chen22} and thus cannot be neglected~\cite{Ayres22} when discussing the physics of the $d$~complex~\cite{NBarisic22}.

\section{Conclusions}

Because our analysis extends over a very broad energy window of 2~eV, across the whole phase diagram of cuprates in all doping regimes (insulator, pseudogap, strange metal, Fermi-liquid), and in both the normal and superconducting states, we are able to characterize the electronic responses of cuprates in full. We have demonstrated that the optical conductivity response is fully consistent with transport measurements, where the low-frequency part corresponds to a FL. By clarifying the important deviation of the scaling parameter from $\pp=2$, we demonstrated perfect Fermi-liquid scaling ($\pp=2$) for the itinerant subsystem in cuprates and other compounds with similar behavior of $\pp$. The experimental demonstration that the optical scattering rate is compound- and doping-independent [even the total one, Figs. \ref{Fig:Hg1201OptCond}(f) and \ref{Fig:Bi2212OptCond}(e)] is direct proof that cuprate physics is not controlled by a QCP: there is no critical behavior in the itinerant subsystem anywhere in the phase diagram, only the gradual crossover of the carrier density from $p$ to $1+p$ as the localized subsystem delocalizes. Essentially, this observation alone lucidly demonstrates that all the approaches based on analyses of scattering rates that neglect the change in carrier density associated with pseudogap evolution (non-Fermi-Liquid, Ioffe-Regel limited, bad metal, quantum critical, strange metal, marginal Fermi liquid, Planckian, etc.) are inappropriate for cuprates.

Our analysis does not require any fitting, because it is fully constrained by the experimental facts determined by independent measurements, namely DC transport. Thus, the decomposition obtained is unambiguous and naturally reveals physical signatures of secondary, localized, electronic subsystems. The high-energy (MIR and higher) part provides valuable insights into the physics of the localized charge, which is responsible for the two principal sources of the cuprates' strangeness, the pseudogap, and the superconducting mechanism itself. These are unequivocally revealed by the spectral weight shifts with temperature and doping, from the high-energy sector to MIR and FL, and finally from FL to SC. These shifts provide important insights into the delocalization process as well.
 
Finally, we have extended our analysis to other compounds, making our conclusions relevant to the wide field of conducting ionic compounds. A clear operative distinction between coherency, incoherency, and (pseudo)gapping is essential for understanding all these materials.


\begin{acknowledgments}
N.B. thanks S. Benhabib for the fruitful discussions. C.M.N.K thanks D. Narayanappa for his input on analysis tools. The mercury cuprate sample was provided by M. Greven. The work at the TU Wien was supported by the European Research Council (ERC Consolidator Grant No. 725521), while the work at the University of Zagreb was supported by project CeNIKS co-financed by the Croatian Government and the European Union through the European Regional Development Fund-Competitiveness and Cohesion Operational Programme (Grant No. KK.01.1.1.02.0013) and by the Croatian Science Foundation under Project No. IP-2018-01-7828. The work at AGH University of Science and Technology was supported by the National Science Centre, Poland, grant OPUS: Grant No. UMO-2021/41/B/ST3/03454, the Polish National Agency for Academic Exchange under 'Polish Returns 2019' Programme: PPN/PPO/2019/1/00014, and the subsidy of the Ministry of Science and Higher Education of Poland. OSB acknowledges the support of the QuantiXLie Center of Excellence, a project co-financed by the Croatian Government and European Union through the European Regional Development Fund - the Competitiveness and Cohesion Operational Programme (Grant KK.01.1.1.01.0004). Work at Brookhaven National Laboratory was supported by the Office of Science, U.S. Department of Energy under Contract No. DE-SC0012704. CMNK acknowledges funding by a QuantEmX grant from ICAM and the Gordon and Betty Moore Foundation through Grant No. GBMF5305. 
\end{acknowledgments}


\begin{appendix}


\section{Materials and Methods}


\subsection{Optical conductivity measurements\label{AppA}}
The temperature-dependent optical conductivity $R(\omega, T)$ of Hg1201 (in the {\it{ab}}-plane) was measured over a wide frequency range from 60 to 22 000~cm$^{-1}$ (7.4~meV to 2.72~eV) on an infrared Fourier transform spectrometer (Bruker Vertex 80v). The reflectance of the sample (\Rs) is compared to the reflectance of an aluminum reference mirror (\Rm). To correct for the sample size and any irregularities in the surface, and to eliminate the effects of the reference mirror, the sample was coated with gold {\it{in-situ}} and the measurements were repeated on the gold-coated sample (\Rgs). The effects of the reference mirror can be removed by dividing the two ratios, 
\begin{equation}
    \left( \frac{\Rs}{\Rm} \right)\left( \frac{\Rgs}{\Rm} \right)^{-1} =
    \frac{\Rs}{\Rgs}
\end{equation}
which yields the reflectance of the sample with respect to gold. The reflectance can be subsequently corrected by multiplying the ratio by the reflectance of gold to yield the absolute reflectance of the sample; silver is typically used in the visible region. The low-frequency reflectance was extrapolated toward zero frequency through the use of the Hagen-Rubens approximation, $R(\omega )\propto 1-\sqrt{\omega }$, fitted to the lowest-frequency measured reflectance data. The high-frequency reflectance was extrapolated by using x-ray scattering functions as described in Ref.~\cite{Tanner15a}. The complex optical conductivity $\sigma (\omega )$ is extracted from $R(\omega )$ by Kramers-Kronig transformation~\cite{Dressel02}. 


\subsection{Optical response of Fermi liquids}
It is shown in Ref.~\cite{Berthod13} that the FL contribution to the dynamical conductivity, associated with itinerant charges, may be expressed  in terms of two dimensionless variables,
\begin{equation}
    \sigma (\omega, T)=\sigma_{dc}\;F\left(\frac{\tilde\omega }{\tilde T},\frac{\tilde\omega }{\tilde T^2}\right)
    \label{AP01gr}
\end{equation}
with $\tilde\omega=\hbar\omega/\pp\pi k_BT_0$ and $\tilde T=T/T_0$. Here, $T_0$ is characterized solely by the quasiparticle lifetime $\tau_{\mathrm{qp}}$ at zero frequency, as discussed below in relation to Eq.~(\ref{eqtau}). For the FL contribution to optical conductivity, discussed in the main text, $\pp$ is fixed, $\pp=2$.

To obtain Eq.~(\ref{AP01gr}), it is assumed that the single-particle self-energy of itinerant charges is momentum-independent, i.e., local, $\Sigma (\vec k, \omega,T)= \Sigma (\omega,T)$. As a consequence of this locality, a significant simplification occurs when $\sigma (\omega,T)$ is calculated. For low frequencies and temperatures that characterize the FL contribution to $\sigma (\omega,T)$,  the leading contributions in $\omega$ and $T$ to the real and imaginary parts of the momentum-independent self-energy are given by
%
\begin{multline}
    \Sigma (\omega, T)=\left(1-Z^{-1}\right)\omega- \\
    \frac{2i}{Z\pp\pi k_BT_0}\left[\omega^2+(\pp\pi k_BT/2)^2\right]\;
    \label{AP02gr}
\end{multline}
%
where $\pp \equiv 2$ for a FL, and $Z$ is the quasi-particle spectral weight, which becomes suppressed by the interaction, $Z\leq1$. Furthermore, when the self-energy $\Sigma (\omega )$ is local, $Z$ defines the renormalization of the Fermi velocity as well, $Z=v_F^*/v_F$. The quasiparticle lifetime $\tau_{\mathrm{qp}}$ is defined by the imaginary part of $\Sigma (\omega, T)$ at zero-frequency, being given by
\begin{equation}
    \hbar/\tau_{\mathrm{qp}}=-2Z{~\mathrm{Im}}\left[\Sigma (0,T)\right]=\pp\pi
    k_BT_0\left(\frac{T}{T_0}\right)^2\;
    \label{eqtau}
\end{equation}
From Eq.~(\ref{AP02gr}), one may obtain the function $F$ in Eq.~(\ref{AP01gr}) in a closed form,
%
\begin{multline}
    F(x,y) = \frac{6i}{\pi^2} \frac{1}{x~r(x,y)} \left\{\psi
    \left(\frac{1}{2}\left[1+r(x,y)-ix \right]\right)\ \right. \\ \left. - ~\psi \left(\frac{1}{2}\left[1+r(x,y)+ix \right ]
    \right)\ \right\}
    \label{eqn:ScalingFunction}
\end{multline}
%
with $r(x,y) = \sqrt{1+x^2-iy}$, and $\psi$ being the digamma function, defined by
\begin{equation}
    \centering
    \psi (z) = \lim_{M \rightarrow \infty} \left[\ln M - \sum_{n=0}^{M}
    \frac{1}{n+z} \right]
    \label{eqn:digamma}
\end{equation}
Notably, the latter two expressions simplify any procedure to fit experimental data. To compare with the experiment, one just needs to evaluate Eq.~(\ref{eqn:ScalingFunction}) along with Eq.~(\ref{eqn:digamma}), and take the real part of $\sigma (\omega,T)$ in Eq.~(\ref{AP01gr}). At all times, the parameters $\sigdc$ and $\tau_{qp}$ necessary to evaluate Eq.~(\ref{AP01gr}) are obtained directly from the transport coefficients in this work, with no adjustable parameters whatsoever.

It is important to emphasize that in our modeling  $\sigdc$ and $\tau_{\mathrm{qp}}$ is used to characterize the optical response of itinerant charges only. The cleanest limit, regarding itinerant charges, is the overdoped regime, where the Fermi-surface is circular exhibiting Fermi-liquid behavior, with $\neff=1+p$. There, the use of the isotropic scattering rate is appropriate and thus the reason to use the local self-energy. As explained in the main text, the density of these charges \neff\ decreases with doping resulting in the nodal arcs that correspond to  $\neff=p$ in the underdoped regime. The main message of the universality of the Hall mobility, confirmed now by optical conductivity, is that the Fermi-velocity/effective mass and $\tau_{\mathrm{qp}}$ remain essentially  unchanged on the ungapped parts of the FS. Surely, $\sigdc$ involves the (ungapped) parts of the FS with significant electron spectral weight only, for which (consistently with experiment) we keep the same form of the self-energy in Eq.~(\ref{AP01gr}) as in the overdoped regime. That is, because the optical conductivity is proportional to the average squared group velocity (the transport function in Ref.~\cite{Berthod13}), involving integration only along the ungapped parts of the FS, the knowledge of the self-energy for these parts is sufficient to derive the behavior of $\sigma(\omega,T)$. The remaining parts of the FS are gapped and thus give no contribution from transport to the optical conductivity.


\subsection{Memory-function formalism}
\begin{figure}[!htb]
    \centering
    \includegraphics[scale=0.5]{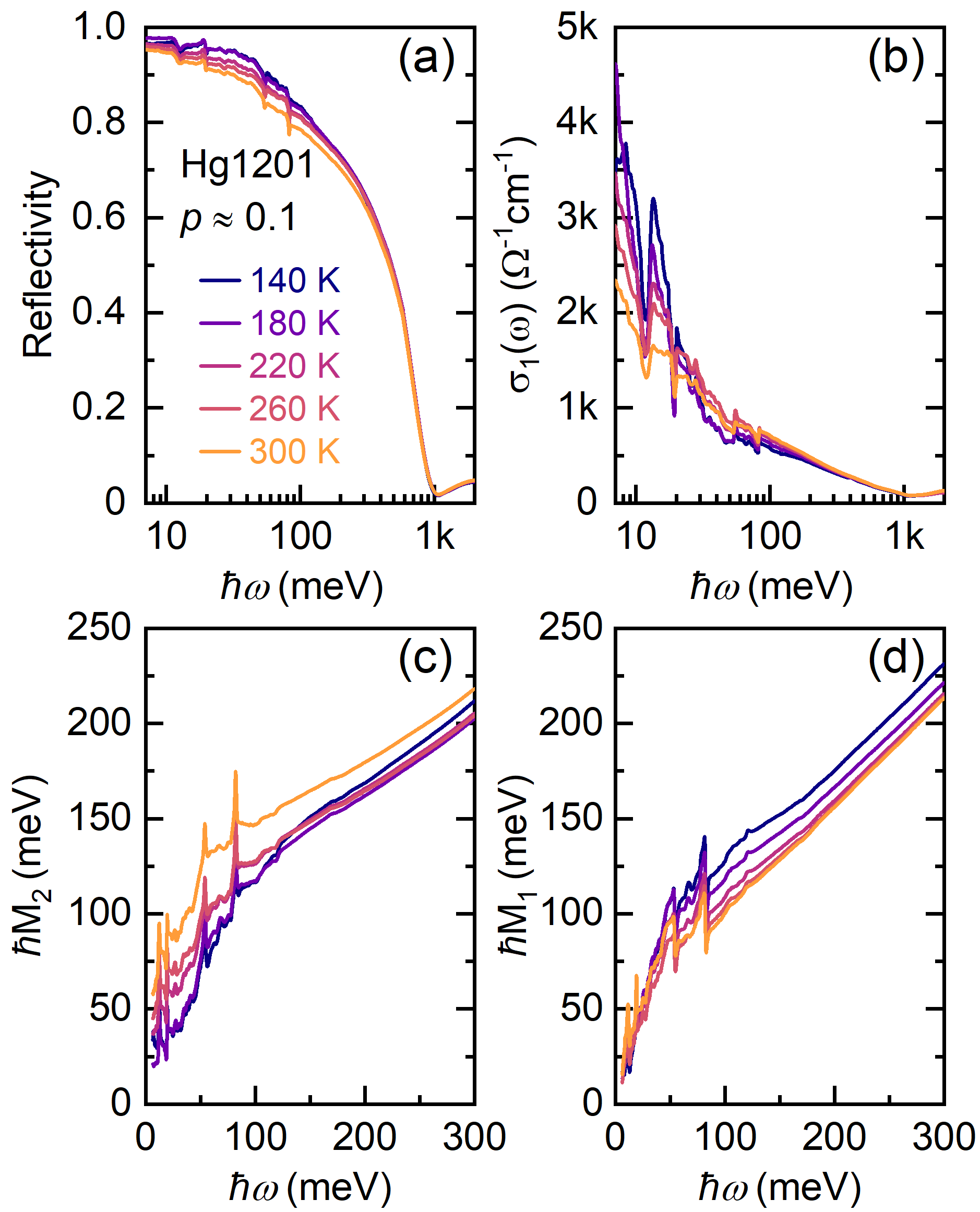}
    \caption{(a) Measured optical reflectivity of underdoped Hg1201 at few selected temperatures. (b) The real part of the optical conductivity obtained from the measured optical reflectivity. (c) Imaginary and (d) real parts of the memory function as a function of $\hbar \omega$.}
    \label{suppFig:Hg1201-Optical-Properties}
\end{figure}

\begin{figure}[!htb]
    \centering
    \includegraphics[scale=0.5]{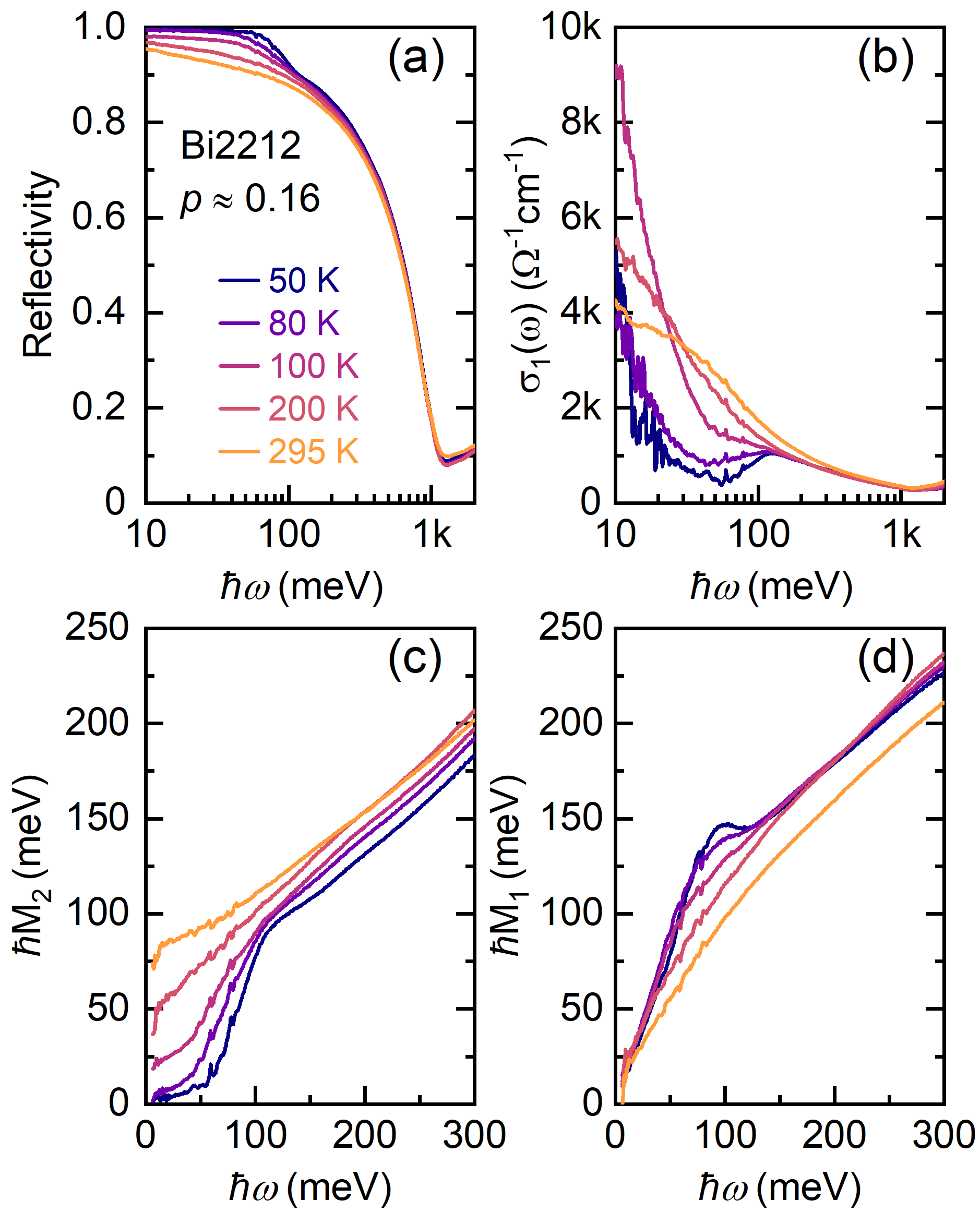}
    \caption{(a) Measured optical reflectivity of optimally doped Bi2212 at few selected temperatures. (b) The real part of the optical conductivity obtained from the measured optical reflectivity. (c) Imaginary and (d) real parts of the memory function as a function of $\hbar \omega$.}
    \label{suppFig:Bi2212-Optical-Properties}
\end{figure}

The optical response of metals is commonly given by the Drude formula~\cite{Drude1900},
\begin{equation}
    \centering
    \sigma (\omega) =  \frac{\sigdc}{1 - i\omega \tau_D}
    \label{eqn:Drude}
\end{equation}
where $\sigma (\omega) \equiv \sigma_1 (\omega) + i\sigma_2 (\omega)$ is the complex optical conductivity, $\sigdc = \sigma (\omega \rightarrow 0) = ne^2\tau_D/m^\ast$ is dc conductivity, and $\tau_D=1/\Gamma_D$ is the frequency-independent scattering time, governing the relaxation of the current. Interaction-related relaxation processes induce a frequency dependence in the scattering time. This leads to a more general expression for the conductivity, given in terms of the complex memory function, $M(\omega) = M_1(\omega)+iM_2(\omega)$~\cite{Gotze72}:
\begin{equation}
    \centering
    \sigma (\omega) = \frac{i\epsilon_0\omega^2_p}{\omega + M(\omega)}
    \label{eqn:DrudeMemory}
\end{equation}
where $\epsilon_0$ is the permittivity of free space, and the plasma frequency is defined from the {\it{f}}-sum rule over the whole frequency range,
\begin{equation}
    \centering
    \epsilon_0\omega^2_p = \frac{2}{\pi }\int_{0}^{\infty}\sigma_1(\omega)d\omega
    \label{eqn:PlasmaFreq}
\end{equation}
Thus, the imaginary part of the memory function may be determined from the optical conductivity $\sigma (\omega)$ data,
\begin{equation}
    \centering
    M_2(\omega )=\mathrm{Re}
    \left[\frac{\epsilon_0\omega^2_p}{\sigma(\omega)}\right]
    \label{eqn:MemoryImaginary}
\end{equation}
Equation~\eqref{eqn:Drude} can be written in a form analogous to the Drude expression,
\begin{equation}
    \centering
    \sigma (\omega)=\epsilon_0\omega^2_p ~\frac{g(\omega)}{-i\omega+1/\tau
    (\omega)}
    \label{eqn:DrudeGeneral}
\end{equation}
with
\begin{equation}
    \centering
    g(\omega)=\left[1+\frac{M_1(\omega)}{\omega}\right]^{-1},\quad\frac{1}{\tau (\omega)} = g(\omega)M_2(\omega)
    \label{eqn:Domega}
\end{equation}
When $g(\omega )$ is nearly constant at low frequencies, the scaling properties of the optical scattering rate $1/\tau (\omega )$ are given by $M_2(\omega)$ only. In particular, for $\pp=2$ the characteristic of the FL is the scaling collapse of $M_2(\omega)$ as a function of $\xi^2$ defined by
Eq.~(\ref{eqn:xisqr}). The steps leading from the experimental reflectivity to the memory function are illustrated in Figs. \ref{suppFig:Hg1201-Optical-Properties} and \ref{suppFig:Bi2212-Optical-Properties} for Hg1201 and Bi2212, respectively.

\section{Optical response in three exemplary systems}
\label{sec:OptResp_Appendx}


\subsection{Systems with a FL contribution only}

\begin{figure}[!htb]
    \centering
    \includegraphics[scale=0.5]{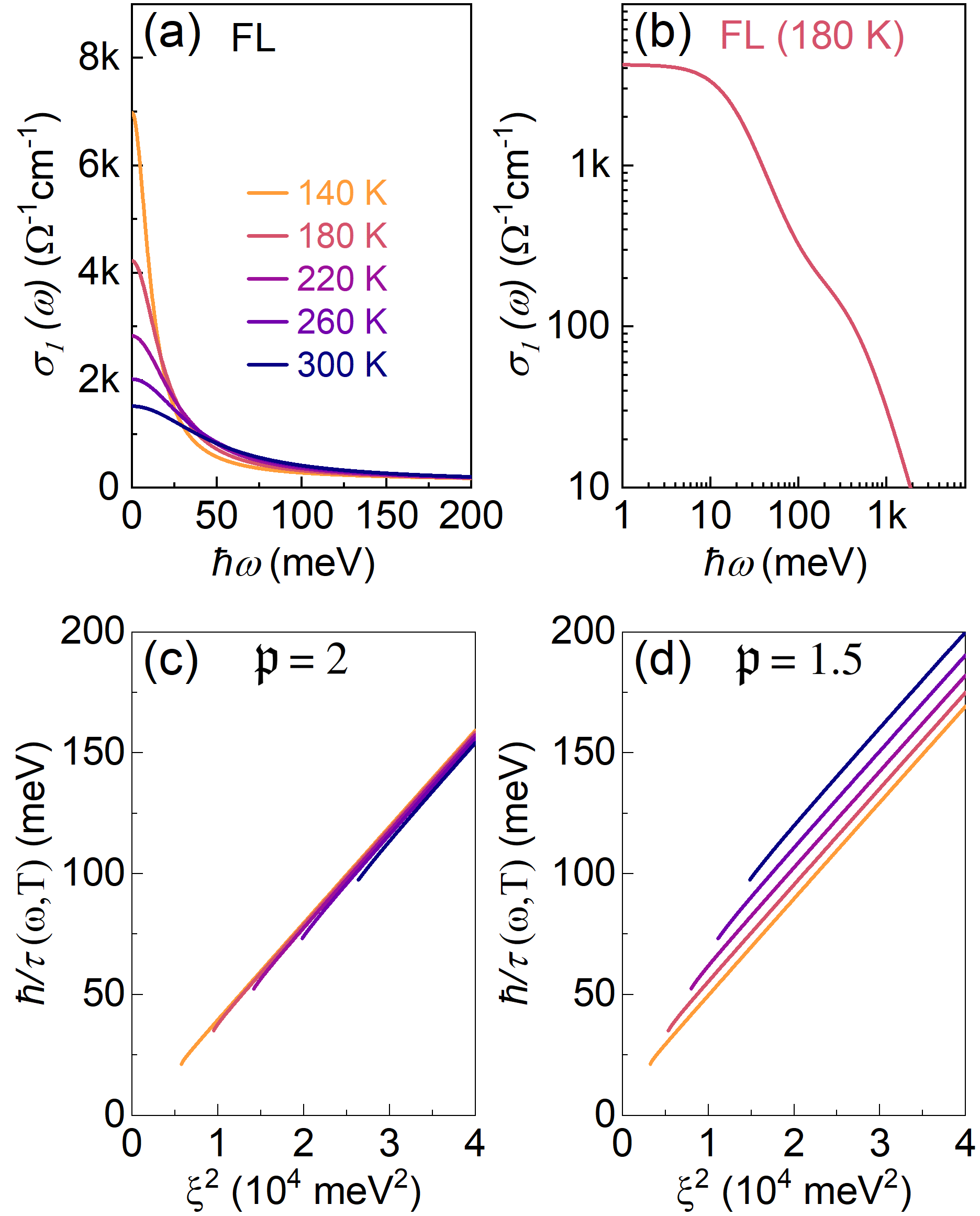}
    \caption{(a) Calculation of itinerant charge (FL) contributions to the optical conductivity for selected temperatures. (b) FL contribution from (a) at a given temperature (180 K) is shown in log-log scale. (c), (d) Scaling behavior of the FL optical scattering rate for the parameter $\pp=2$ and $1.5$ demonstrating that the scaling is adequately sensitive to the value of \pp.}
    \label{suppFig:Ext-Drude-Sim_FL}
\end{figure}

For the calculation of pure FL-like optical conductivity shown in Fig.~\ref{suppFig:Ext-Drude-Sim_FL}, we used resistivity and $\cot{\Theta_H}$ data of Hg1201 at 180~K and generated the corresponding data for other temperatures by using the universal relations that govern transport properties of FL compounds,
\begin{equation}
    \centering
    \begin{aligned}
        \rho (T) = AT^2 \Rightarrow \sigdc(T) &=1/AT^2 \\
        \Gamma_D(T) &= A^\prime T^2
    \end{aligned}
    \label{UniversalRhoGamma}
\end{equation}
where $\Gamma_D(T) = 1/\tau_D$ is calculated from $\cot{\Theta_H}$ using the relation,
\begin{equation}
    \centering
    \cot{\Theta_H} = \frac{m^{\ast}}{e H \tau_D}
    \label{cotH}
\end{equation}

By introducing the simulated dc transport parameters from Eqs.~(\ref{UniversalRhoGamma}) and (\ref{cotH}) into Eq.~(\ref{AP01gr}), we calculated FL optical conductivity at several temperatures presented in
Figs.~\ref{suppFig:Ext-Drude-Sim_FL}(a)-\ref{suppFig:Ext-Drude-Sim_FL}(b). The scaling behavior of $M_2$ was calculated using Eq.~(\ref{eqn:MemoryImaginary}) from the aforementioned simulated optical conductivity data, and tested for several $\pp$ values. As expected for a FL, the scaling collapse occurs for $\pp=2$, as shown in Fig.~\ref{suppFig:Ext-Drude-Sim_FL}(c).

%
%


%
%
\begin{figure}[!htb]
    \centering
    \includegraphics[scale=0.5]{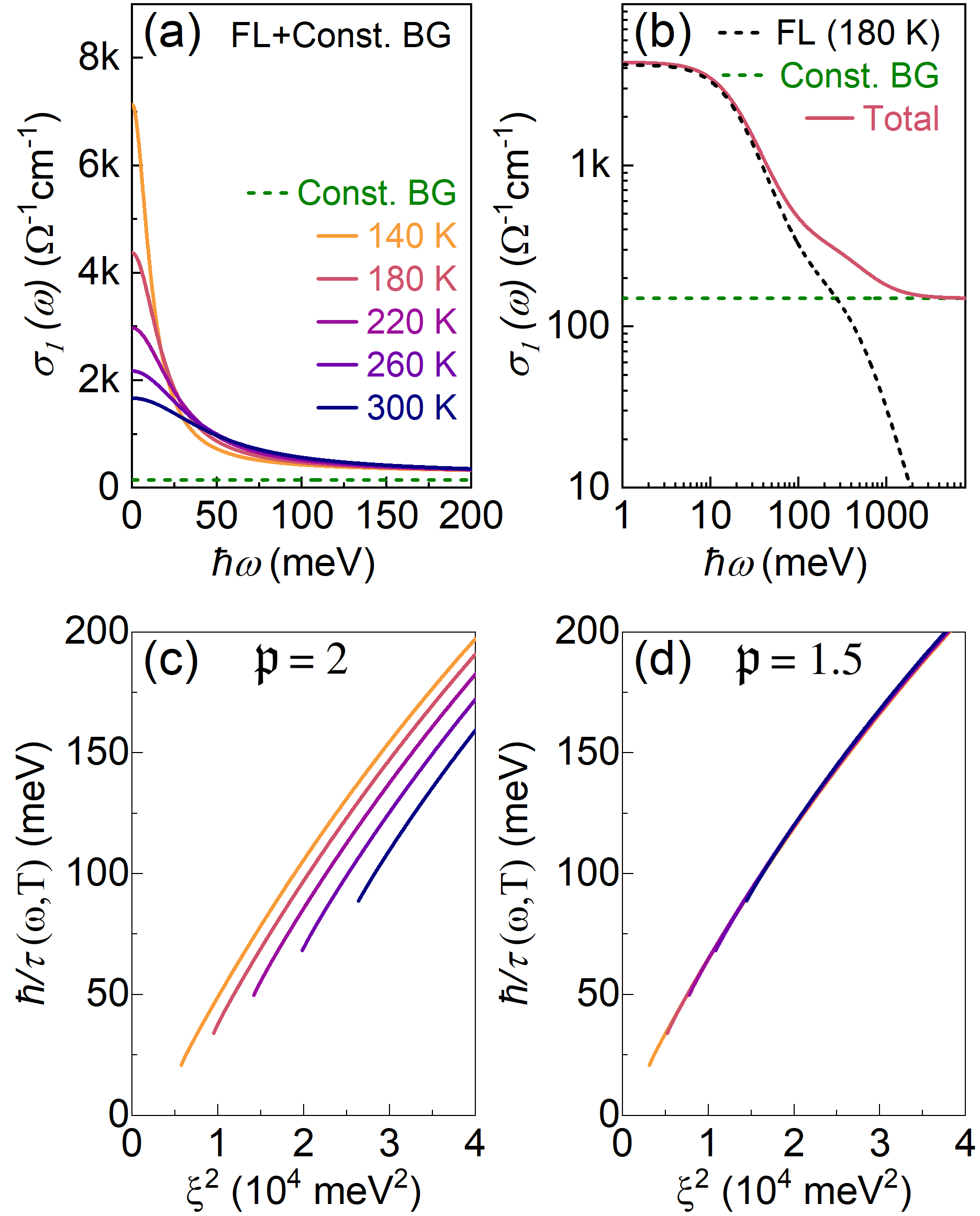}
    \caption{(a) Calculation of total $\sigma_1(\omega,T)$ due to itinerant charges (FL) (as shown in Fig. \ref{suppFig:Ext-Drude-Sim_FL}) plus a small constant background (150~$\Omega^{-1}$cm$^{-1}$) contribution to the optical conductivity at selected temperatures. The constant background used mimics contributions from incoherent scattering effects. (b) Simulated total optical conductivity at a representative temperature (180~K) shown in log-log scale along with FL and Lorentzian components. (c), (d) Scaling behavior of the optical scattering rate for the parameter $\pp=2$ and $\pp=1.5$. A small contribution associated with the incoherency effects shifts the scaling collapse to smaller $\pp$-values ($\pp=1.5$ in this case).}
    \label{suppFig:Ext-Drude-Sim_FL-Incoh}
\end{figure}


\subsection{System with contributions from a FL and additional incoherent scattering effects}
An interesting possibility is that in addition to a FL contribution, part of the response of a system is  incoherent. Such behavior should be expected in systems with a Fermi surface that contains a shallow pocket or a flat band. Once $k_B T$ is of the order of $E_F$ or of the bandwidth, respectively, the corresponding part of the Fermi surface should be smeared out by temperature and consequently become incoherent. Thus the overall response should correspond to a sum of the FL and an incoherent part, which we mimic in optical conductivity by a small constant background [green dashed line in Figs.~\ref{suppFig:Ext-Drude-Sim_FL-Incoh}(a)-\ref{suppFig:Ext-Drude-Sim_FL-Incoh}(b)]. The overall optical scattering rate does not scale with $\pp=2$ any more [Fig.~\ref{suppFig:Ext-Drude-Sim_FL-Incoh}c], though the FL part does; however, it scales nicely with $\pp=1.5$ (for the chosen incoherent
background).

%
%
\begin{figure*}[!htb]
    \centering
    \includegraphics[scale=0.45]{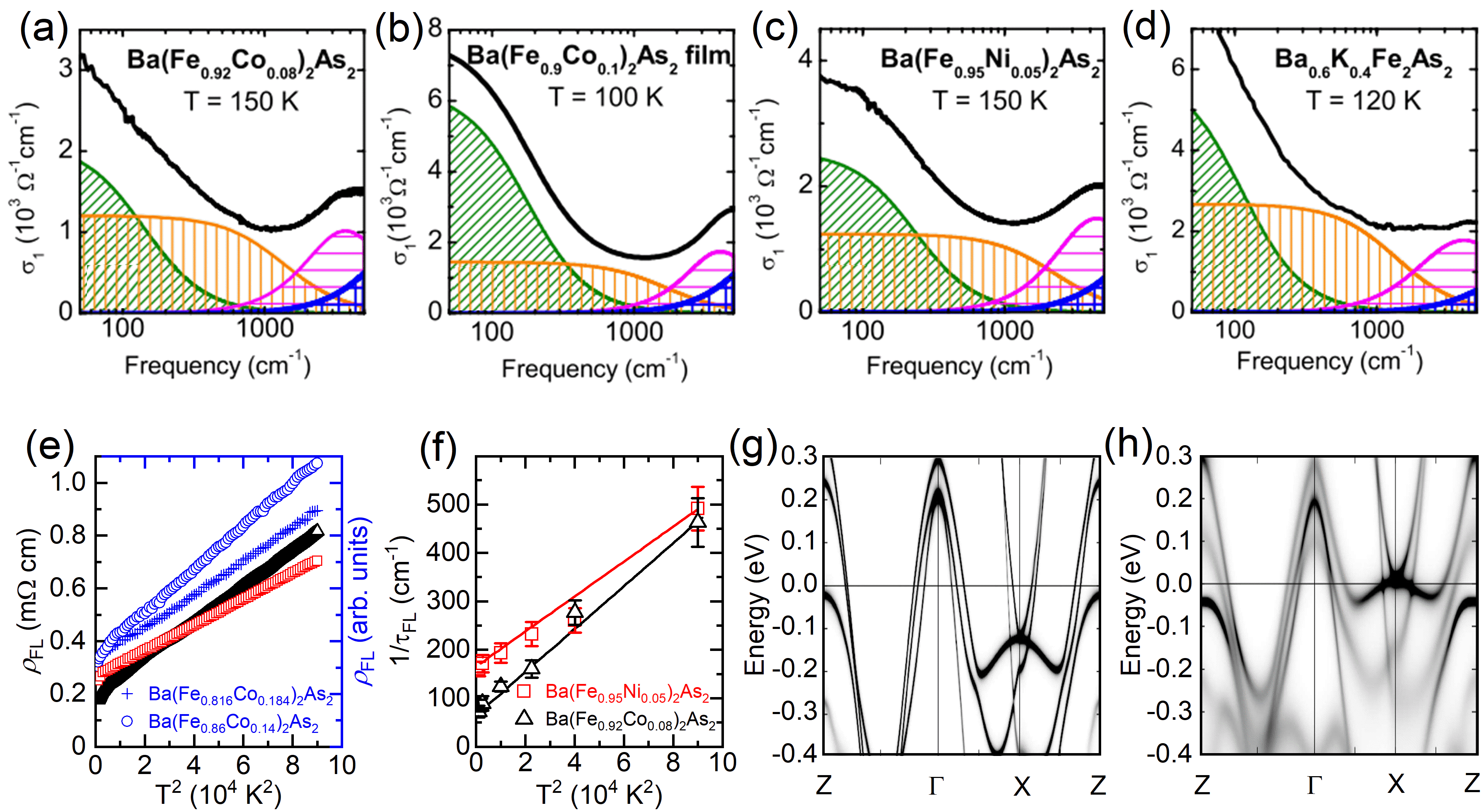}
    \caption{(a)-(d) The optical conductivity response measured for several 122-pnictide superconductors (reproduced from~\cite{NBarisic10}). A two component model was used to fit the optical data in the energy range where only intra-band transitions are expected. The green hashed region corresponds to an itinerant charge contribution, fitted with a Drude term and the broad orange hashed area was used to express the temperature independent contribution attributed to incoherency~\cite{Wu10,NBarisic10}. (e) Temperature dependence of FL part of resistivity plotted as a function of $T^2$ (reproduced from~\cite{NBarisic10}). For a variety of iron pnictides Ba(Fe$_{0.95}$Ni$_{0.05}$)$_{2}$As$_{2}$ (red squares), Ba(Fe$_{0.92}$Co$_{0.08}$)$_{2}$As$_{2}$ (black triangles), Ba(Fe$_{0.86}$Co$_{0.14}$)$_{2}$As$_{2}$ (blue circles) and Ba(Fe$_{0.816}$Co$_{0.184}$)$_{2}$As$_{2}$ (blue pluses) it exhibits a quadratic, FL, temperature dependence all the way up to room temperature. (f) Temperature dependence of the optical scatting rate $1/\tau$ of Ba(Fe$_{1-x}$M$_{x}$)$_{2}$As$_{2}$ obtained from the temperature dependence of the FL-contribution (green hashed area) also exhibits $T^2$ behavior (reproduced from~\cite{NBarisic10}). (g) Band structure of (Ba$_{0.6}$K$_{0.4}$)Fe$_{2}$As$_{2}$ calculated on the basis of LDA and (h) on the basis of LDA+DMFT [reproduced from~\cite{Derondeau17}]. The latter fits the ARPES data well and reveals a flat band pinned to the $E_F$ at X-point.} \vspace{4pt}
    \label{suppFig:Pnictide_OptResARP}
\end{figure*}
%

%
%
\begin{figure*}[!htb]
    \centering
    \includegraphics[scale=0.48]{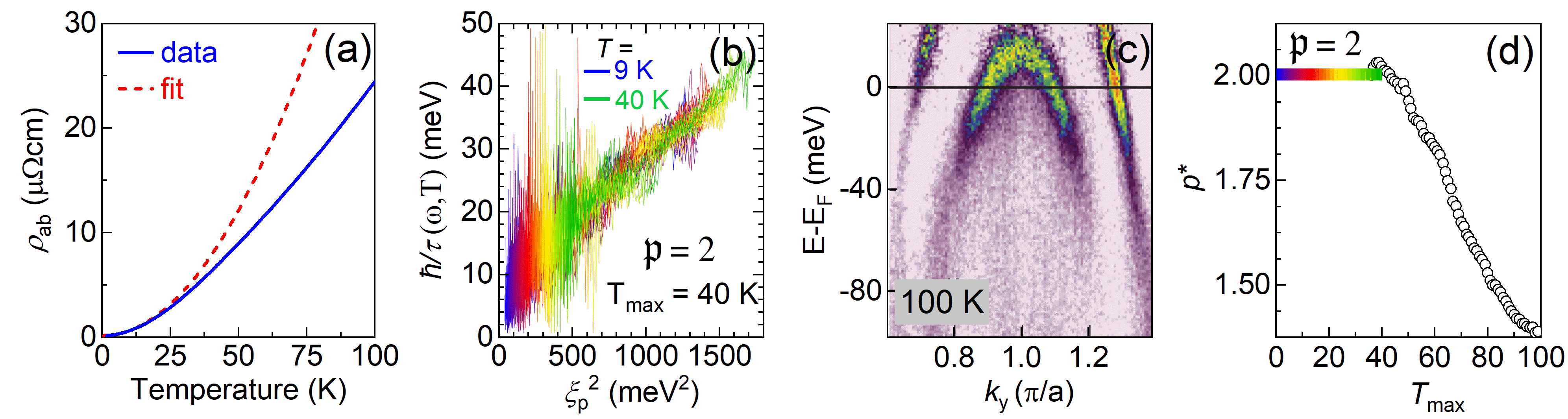}
    \caption{(a) Temperature dependence of resistivity in Sr$_2$RuO$_4$ (blue line)~\cite{Stricker15}. Low temperature fit to $\rho (T) = aT^{2}+\rho_0$ (dashed line). (b) FL scaling of the optical relaxation rate below $T \leq 40~K$ (reproduced from~\cite{Stricker15}). (c) Dispersion map of Sr$_2$RuO$_4$ revealing a 4--5~meV shallow Fermi-pocket (reproduced from~\cite{Burganov16}). (d) Deviation of $\pp$ from 2 for the collapse of scattering rate is observed at $T>40$~K and reaches as low as 1.35 around 100~K (data from~\cite{Stricker14}). The horizontal bar corresponds to $\pp=2$, as obtained from the scaling shown in (b).}
    \label{suppFig:SRO_RhoOptArpes}
\end{figure*}

It is known that the low-energy spectrum of 122-pnictides cannot be modeled with DFT+U alone [Fig.~\ref{suppFig:Pnictide_OptResARP}g], but needs a DMFT correction [Fig.~\ref{suppFig:Pnictide_OptResARP}h]~\cite{Derondeau17}, in sharp contrast with the cuprates. Irrespective of the reasons for this difference~\cite{Sunko20a}, the bare fact that DMFT is required to fit ARPES at the Fermi level indicates that part of the low-energy contribution could be due to non-FL corrections to FL behavior in the conducting subsystem itself. Again, the keyword is ``correction''. And indeed, two electronic subsystems were identified from optical conductivity in 122-pnictides [Figs.~\ref{suppFig:Pnictide_OptResARP}(a)-\ref{suppFig:Pnictide_OptResARP}(d)]~\cite{Wu10,NBarisic10}. One, $\sigma_{\mathrm{incoh}}$ (brown hatched area) corresponds to an incoherent contribution (temperature independent), while the other, $\sigma_{\mathrm{FL}}$ (green hatched area) was identified as a FL. Figures~\ref{suppFig:Pnictide_OptResARP}(a)-\ref{suppFig:Pnictide_OptResARP}d) are a nice example of how this incoherency manifests itself in the overall resistivity. The measured overall resistivity ($\rho_{Tot}$) exhibits a typical ``non-Fermi'' liquid response $\rho \propto AT^n$, with $n<2$. However, an appropriate subtraction of the incoherent response at $\omega=0$ [one just reads it from Figs.~\ref{suppFig:Pnictide_OptResARP}(a)-\ref{suppFig:Pnictide_OptResARP}(d)] from the (measured) $\rho_{\mathrm{Tot}}$  (1/$\rho_{\mathrm{Tot}}=\sigma_{\mathrm{Tot}}=\sigma_{\mathrm{incoh}}+\sigma_{\mathrm{FL}}$) reveals that the resistivity associated with the FL ($\rho_{\mathrm{FL}}=1/\sigma_{\mathrm{FL}}$) exhibits a quadratic temperature dependence from \Tc\ to at least room temperature [Fig.~\ref{suppFig:Pnictide_OptResARP}e]~\cite{Wu10,NBarisic10}. The same temperature dependence is revealed for the optical scattering rate of the FL part from $\sigma_{FL}$ by following the temperature dependence of the half-width at half-maximum [Fig.~\ref{suppFig:Pnictide_OptResARP}f]~\cite{Wu10,NBarisic10}. In contrast, in cuprates, the PG regime is marked by pure $\rho=A_2 T^2$ while across the phase diagram $1/\mu_H=C_2T^2$. \emph{Both eliminate any possibility of an incoherent contribution with a non-zero response at $\omega=0$, because the MIR feature has a gap in the spectrum as a consequence of the PG,} with necessarily zero response at $\omega=0$, as discussed in the main text and below.

\clearpage
%
%

%
%
\begin{figure}[!t]
        \centering
        \includegraphics[scale=0.5]{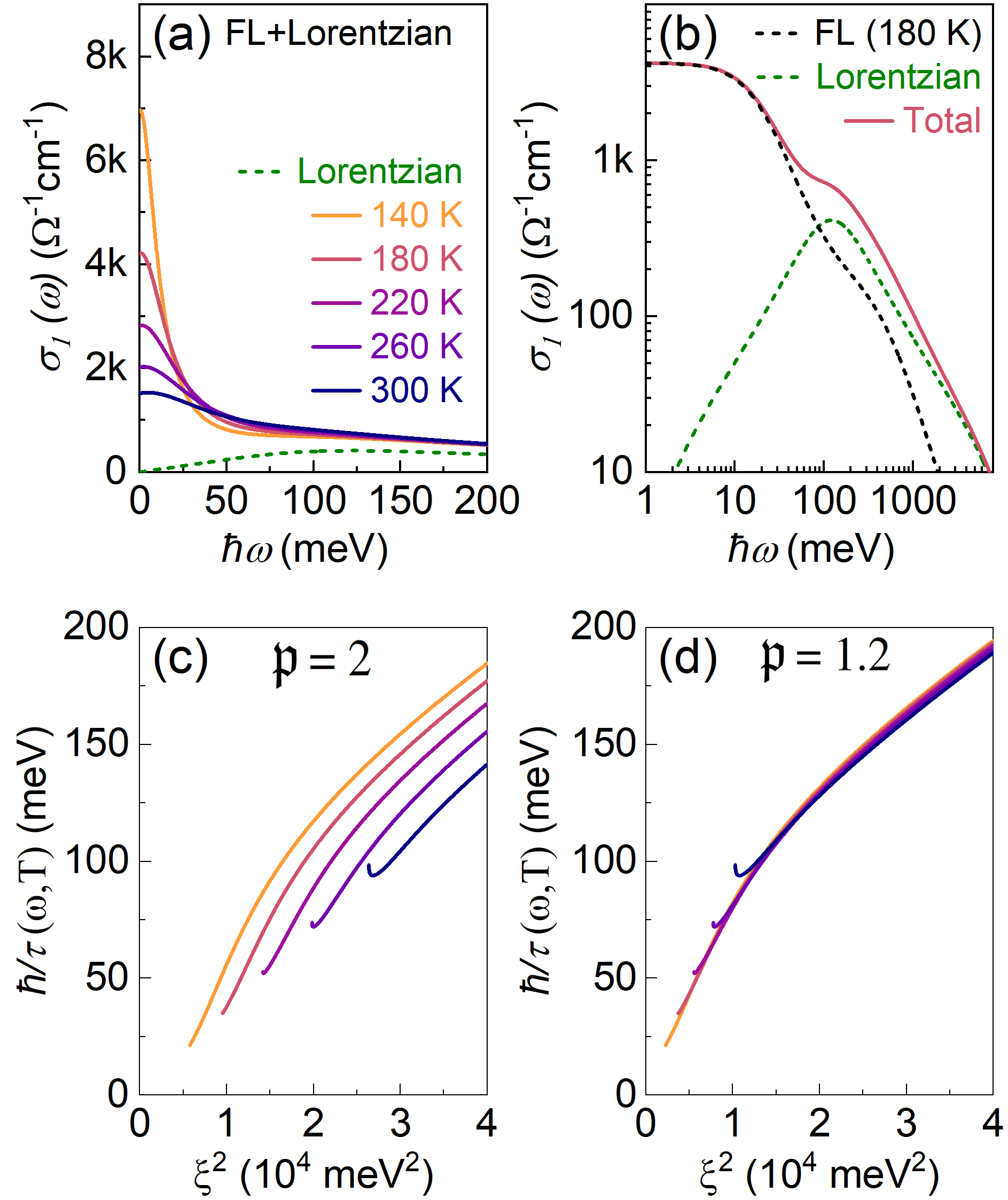}
        \caption{(a) Simulation of total contribution due to itinerant charges (FL) (as shown in Fig. \ref{suppFig:Ext-Drude-Sim_FL}) plus a small Lorentzian-type contribution (centered at $\hbar\omega = 125$~meV) to the optical conductivity at selected temperatures. The Lorentzian mimics the contribution from a gapped system (in cuprates, it can be associated with localized charges). \textbf{(b)} Total optical conductivity at a representative temperature (180~K) shown in log-log scale along with FL and Lorentzian components. (c), (d) Scaling behavior of the optical scattering rate for choices of the parameter $\pp=2$ and $1.2$. A small additional contribution (due to a gap) shifts $\pp$ to considerably smaller values ($\pp = 1.2$ in the case here).}
        \label{suppFig:Ext-Drude-Sim_FL-Gap}
\end{figure}

It is very well established that the dc resistivity of the $4d$ compound Sr$_2$RuO$_4$ (SRO) exhibits $T^2$ temperature dependence below 25~K. Upon further increase of the temperature, the resistivity gradually becomes linear-like (Fig.~\ref{suppFig:SRO_RhoOptArpes}a)~\cite{Mackenzie03, Stricker15}. Recently, it was demonstrated that the optical scattering rate at low temperatures also exhibits pure FL scaling, with $\pp=2$~\cite{Stricker14}. At $40$~K, \pp\ starts to deviate from $2$ and reaches values as low as $1.35$ at $   \sim 100$~K [Figs.~\ref{suppFig:SRO_RhoOptArpes}(b,d)] ~\cite{Stricker14}. Notably, the ARPES measurements show a Fermi-pocket of the order 4--5~meV \cite{Burganov16}. In our view, it is to be expected that once $k_B T \approx E_F$ of the pocket (energy that corresponds well to 40~K) this part of the Fermi-surface becomes gradually incoherent. As in the 122-pnictides, this decoherence should manifest itself in the resistivity as a deviation from $T^2$. In optical conductivity, it appears as a constant background which increases with temperature. The incoherent component leads in turn to a deviation of $\pp$ from 2, in agreement with the analysis shown in Fig.~\ref{suppFig:Ext-Drude-Sim_FL-Incoh}.


%
%
\begin{figure}[!t]
    \centering
    \includegraphics[scale=0.5]{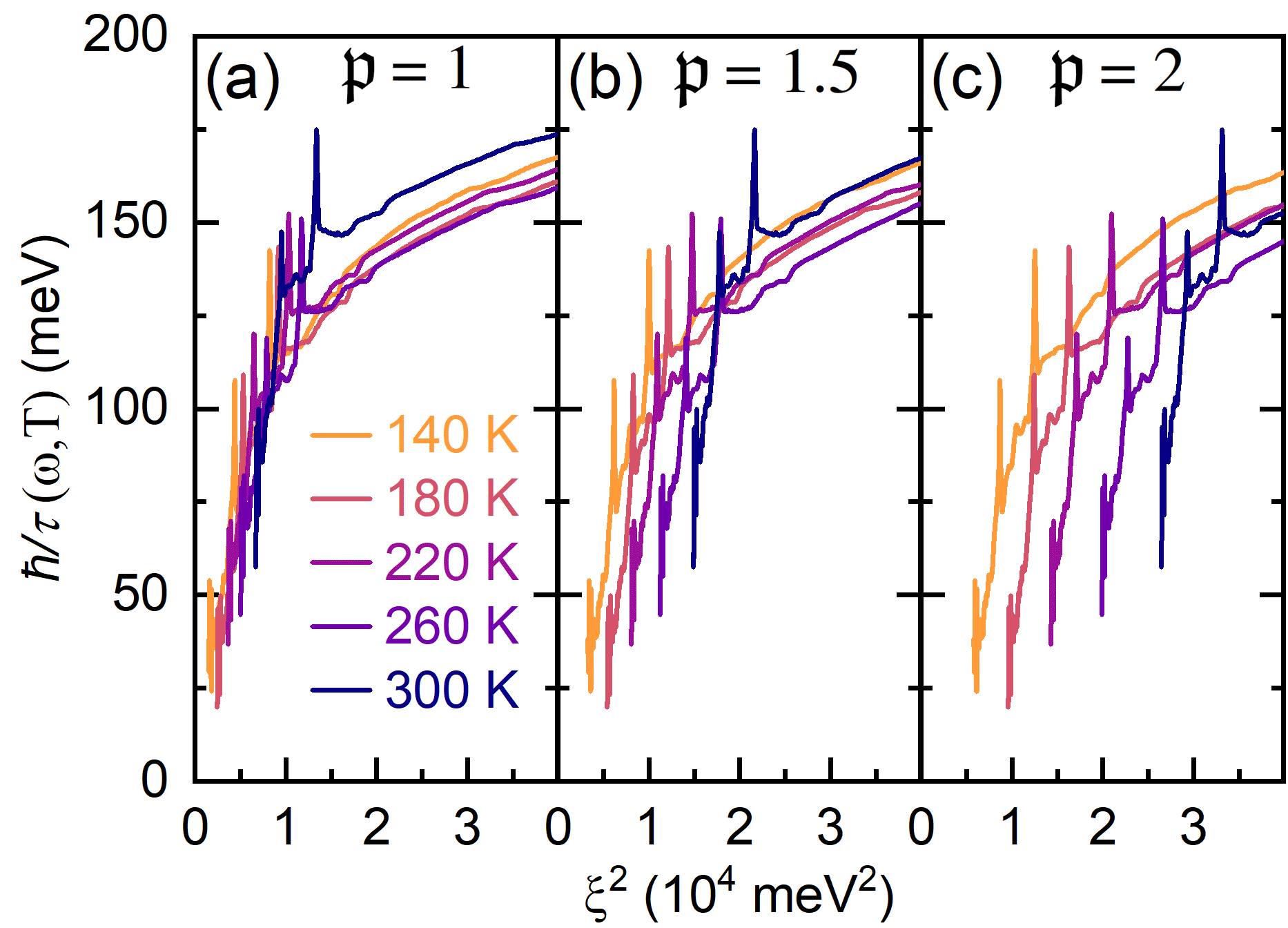}
    \caption{The scattering rate of underdoped ($p\approx 0.1$) Hg1201 for scaling parameter, (a) $\pp=1$, (b) $\pp=1.5$, and (c) $\pp=2$. The overall scattering rates are calculated from the experimental optical conductivity data presented in Fig.~\ref{suppFig:Hg1201-Optical-Properties}.
    }
    \label{suppFig:HG1201_Scaling}
\end{figure}
%

%
%
\begin{figure}[!b]
    \centering
    \includegraphics[scale=0.5]{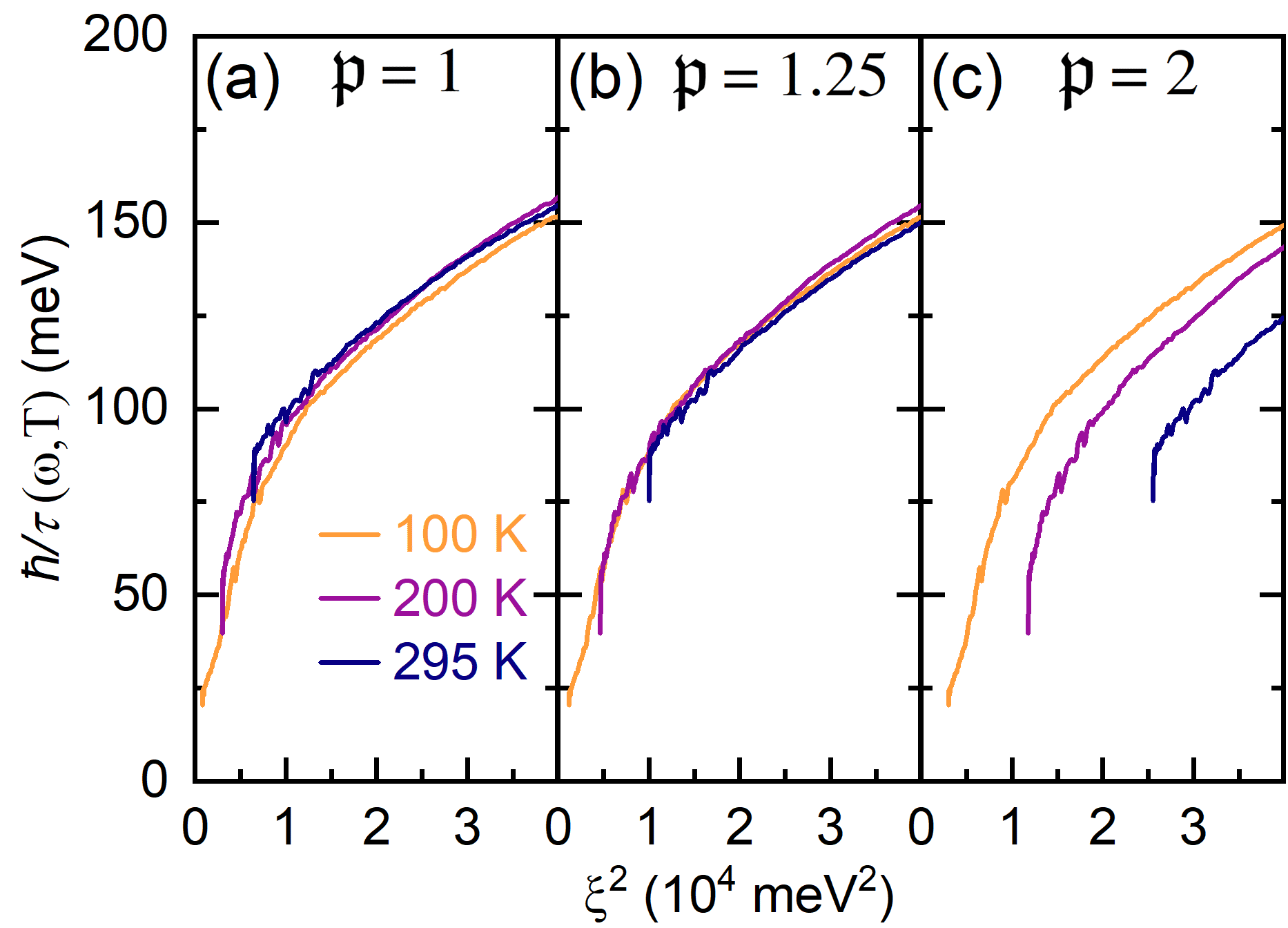}
    \caption{The scattering rate of optimally doped ($p\approx 0.16$) Bi2212 for scaling parameter, (a) $\pp=1$, (b) $\pp=1.25$, and (c) $\pp=2$. The overall scattering rates are calculated from the experimental optical conductivity data presented in Figs.~\ref{suppFig:Bi2212-Optical-Properties}.
    }
    \label{suppFig:Bi2212_Scaling}
\end{figure}
\subsection{System with contributions from both a FL subsystem and a gapped subsystem}

We mimic the gap-like contributions to optical conductivity by using a Lorentzian, for simplicity [Figs.~\ref{suppFig:Ext-Drude-Sim_FL-Gap}(a)-\ref{suppFig:Ext-Drude-Sim_FL-Gap}(b)] in addition to the pure FL contributions [shown in Fig.~\ref{suppFig:Ext-Drude-Sim_FL}a]. With this additional contribution we can see, again, that the overall optical scattering rate, does not scale with $\pp=2$ anymore [Fig.~\ref{suppFig:Ext-Drude-Sim_FL-Gap}c]. Now the scaling is obtained for $\pp=1.2$ [Fig.~\ref{suppFig:Ext-Drude-Sim_FL-Gap}d]. The simulations shown in Fig.~\ref{suppFig:Ext-Drude-Sim_FL-Gap} are relevant for cuprates, as elaborated in detail in the main text. The compounds of interest were Hg1201 in the underdoped regime and Bi2212 at optimal doping. In both cases, the overall optical scattering rate can be scaled in the FL fashion, but with a \pp\ considerably smaller than 2. In the case of Hg1201, the best fit is obtained by choosing $\pp=1$ (Fig.~\ref{suppFig:HG1201_Scaling}), while a good scaling value for Bi2212 is $\pp\approx1.4$ (Fig.~\ref{suppFig:Bi2212_Scaling}).
\subsection{Optical response and experimental uncertainties of the parameters $C_2$ and \ms}

\begin{figure}[!t]
    \centering
    \includegraphics[clip=true,trim=40 0 0 0, scale=0.35]{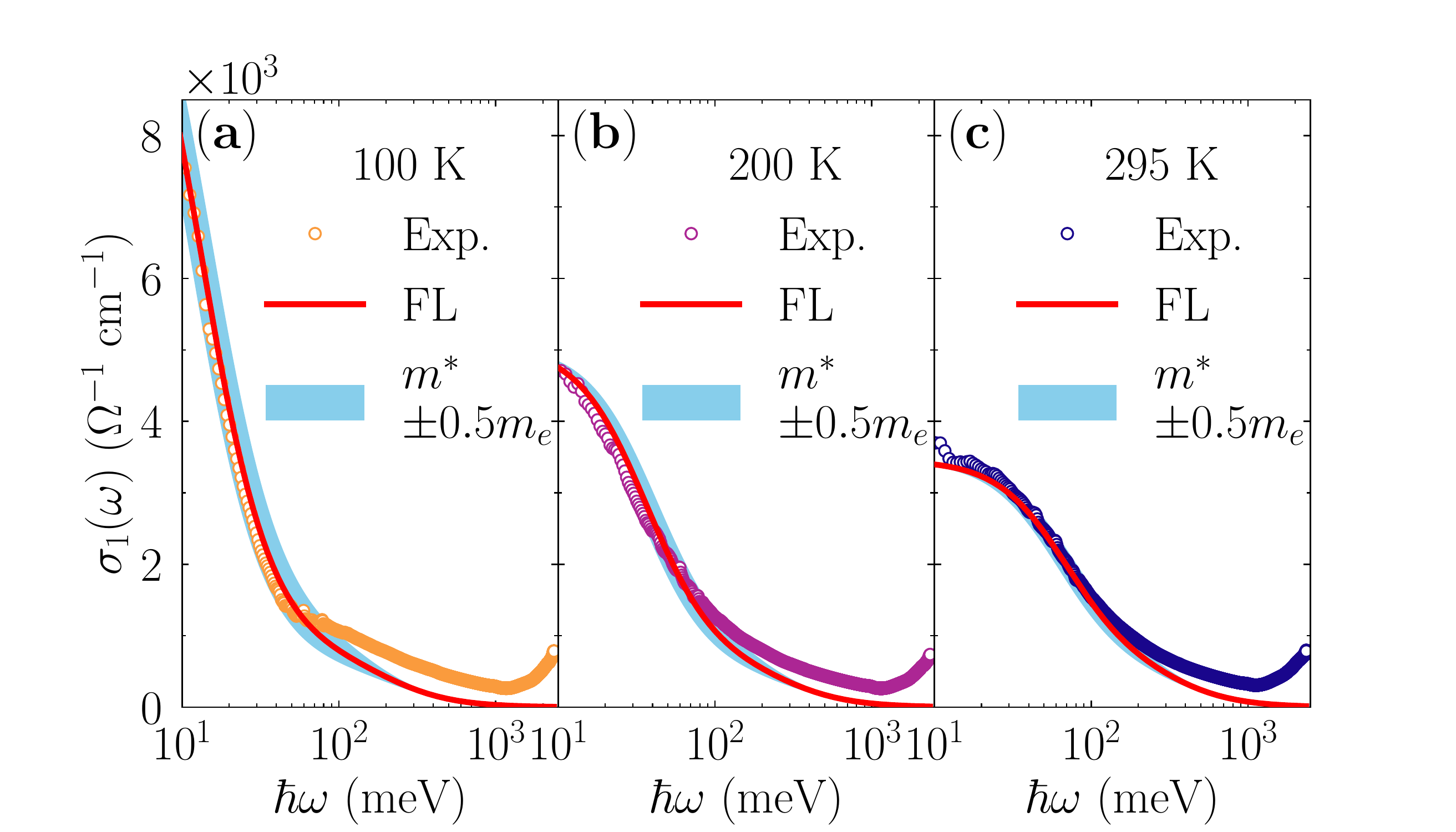}
    \caption{(a--c) As in Fig.~\ref{Fig:Bi2212OptCond} we show the experimental optical conductivity (open circles)~\cite{Tu02} and calculated FL contribution (red line) of optimally doped Bi2212 ($p\approx0.16$) at selected temperatures ($T>T_c$), with $\ms=3\,m_e$, $C_2=0.021\, {\rm T}{\rm K}^{-2}$. The light-blue band corresponds to the FL contribution calculated with $\ms=3\, \pm 0.5 m_e$ (i.e. $\approx 15\%$), which also captures well (essentially within the error bar of the measurement) the low-frequency part of the conductivity spectra for all temperatures (either laying slightly above or slightly below of the experimental points) consistently.}
    \label{suppFig:m*}
\end{figure}

\begin{figure}[!t]
    \centering
    \includegraphics[clip=true,trim=40 0 0 0, scale=0.35]{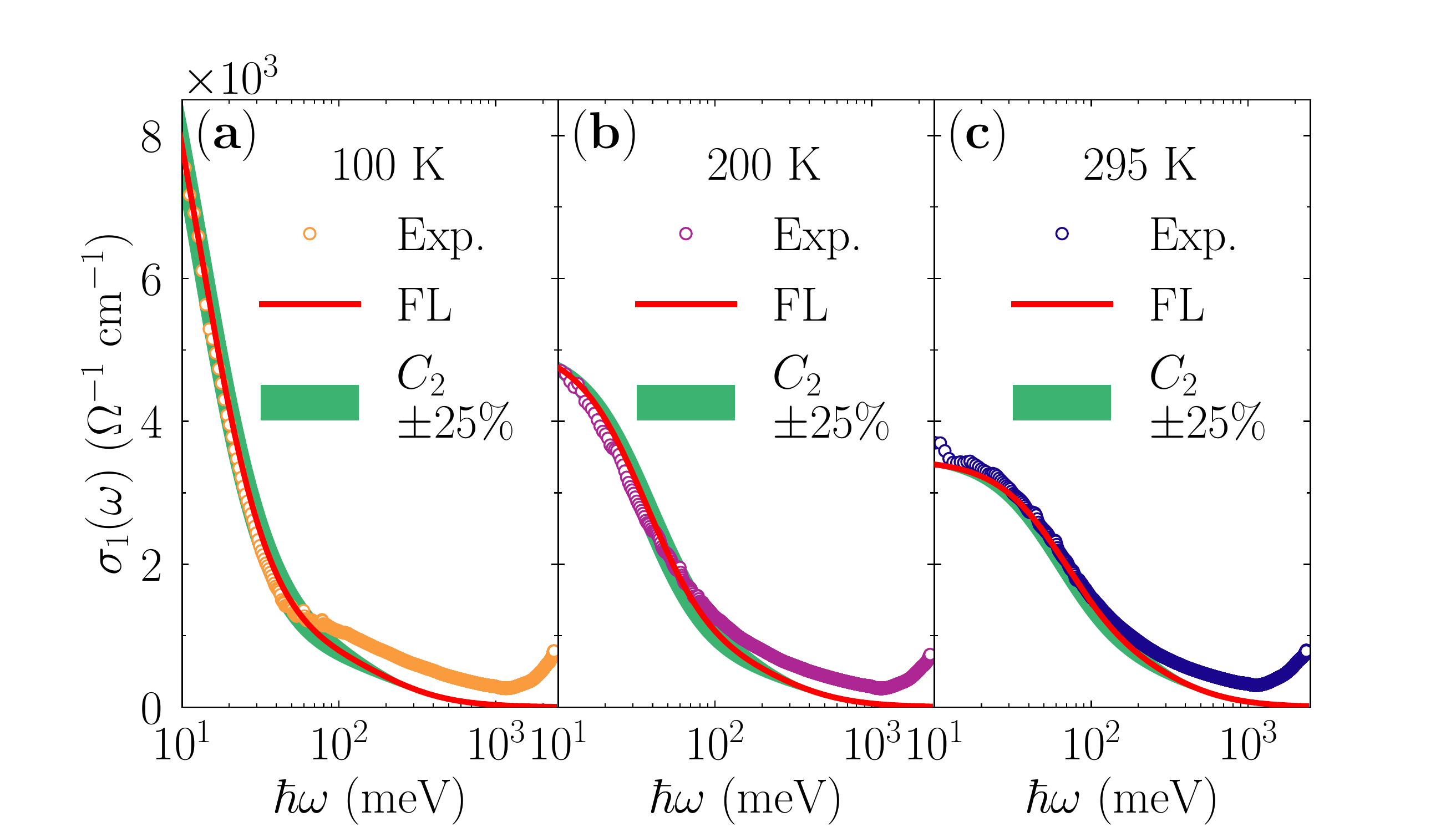}
    \caption{(a--c) As in Figs.~\ref{Fig:Bi2212OptCond} and \ref{suppFig:m*} we show the experimental optical conductivity (open circles)~\cite{Tu02} and calculated FL contribution (red line) of optimally doped Bi2212 (${p\approx0.16}$) at selected temperatures ($T>T_c$), with $\ms=3m_e$, $C_2=0.021\, {\rm T}{\rm K}^{-2}$. The green band corresponds to the FL contribution calculated with $C_2=0.021 \pm 25\%\,{\rm T}{\rm K}^{-2}$~\cite{NBarisic19}, which also well (essentially within the error-bar of the measurement) captures the low-frequency part of the conductivity spectra for all temperatures (either laying slightly above or slightly below of the experimental points) consistently.}
    \label{suppFig:C2}
\end{figure}

Every physical quantity is measured to a certain precision. It is thus expected that the values  $C_2$ and \ms\ vary moderately across different compounds, because they depend on the details of the Fermi surface. The principal aim of our work is to capture the main features of cuprates in zeroth order, providing a solid foundation for addressing all other, often compound-related, issues as well. Nevertheless, it is also important to demonstrate that the proposed approach is not unduly sensitive to the choice of input parameters. This robustness is shown on the example of BSCO, where varying the effective mass by ${\approx}15\%$  (Fig.~\ref{suppFig:m*}) as well as the coefficient $C_2$ by $25\%$  (Fig.~\ref{suppFig:C2}) does not result in any quantitative (not even qualitative) changes and does not affect our understanding of the optical conductivity or the physics of cuprates. Thus, any combination of $C_2$ and \ms\ that captures the data reasonably well within experimentally established limits~\cite{NBarisic19} can be used to calculate the FL response.

%
%
\begin{figure}[!t]
    \centering
    \includegraphics[scale=0.63]{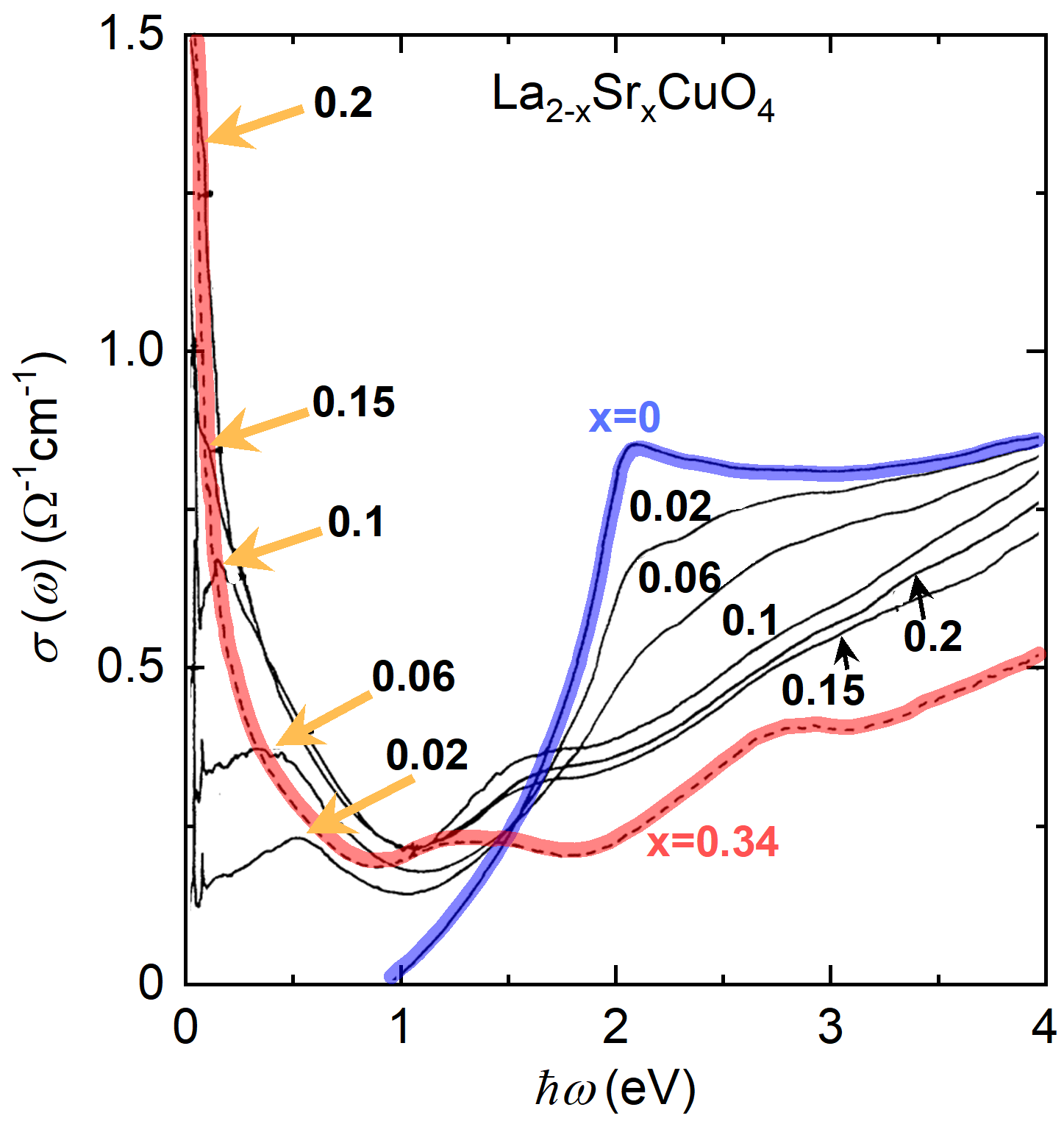}
    \caption{Delocalization of exactly one hole per CuO$_2$ unit can be easily followed from one of the earliest optical spectroscopy measurements performed in LSCO (reproduced from~\cite{Uchida91}). The parent compound (at $x=0$) is a charge-transfer insulator and a clean charge-transfer gap is observed in optical conductivity. Immediately upon doping, a mid-infrared feature appears. Its maxima for various dopings are indicated by yellow arrows. The most important observation, in the context of this paper, is that the mid-infrared feature transfers the spectral weight of the one localized hole with increasing doping from high energies (above the charge-transfer gap) to the coherent (FL) peak in the strongly overdoped regime. 
    }
    \label{suppFig:LSCO_OptResp}
\end{figure}
%

%
%
\begin{figure}[!t]
    \centering
    \includegraphics[scale=0.53]{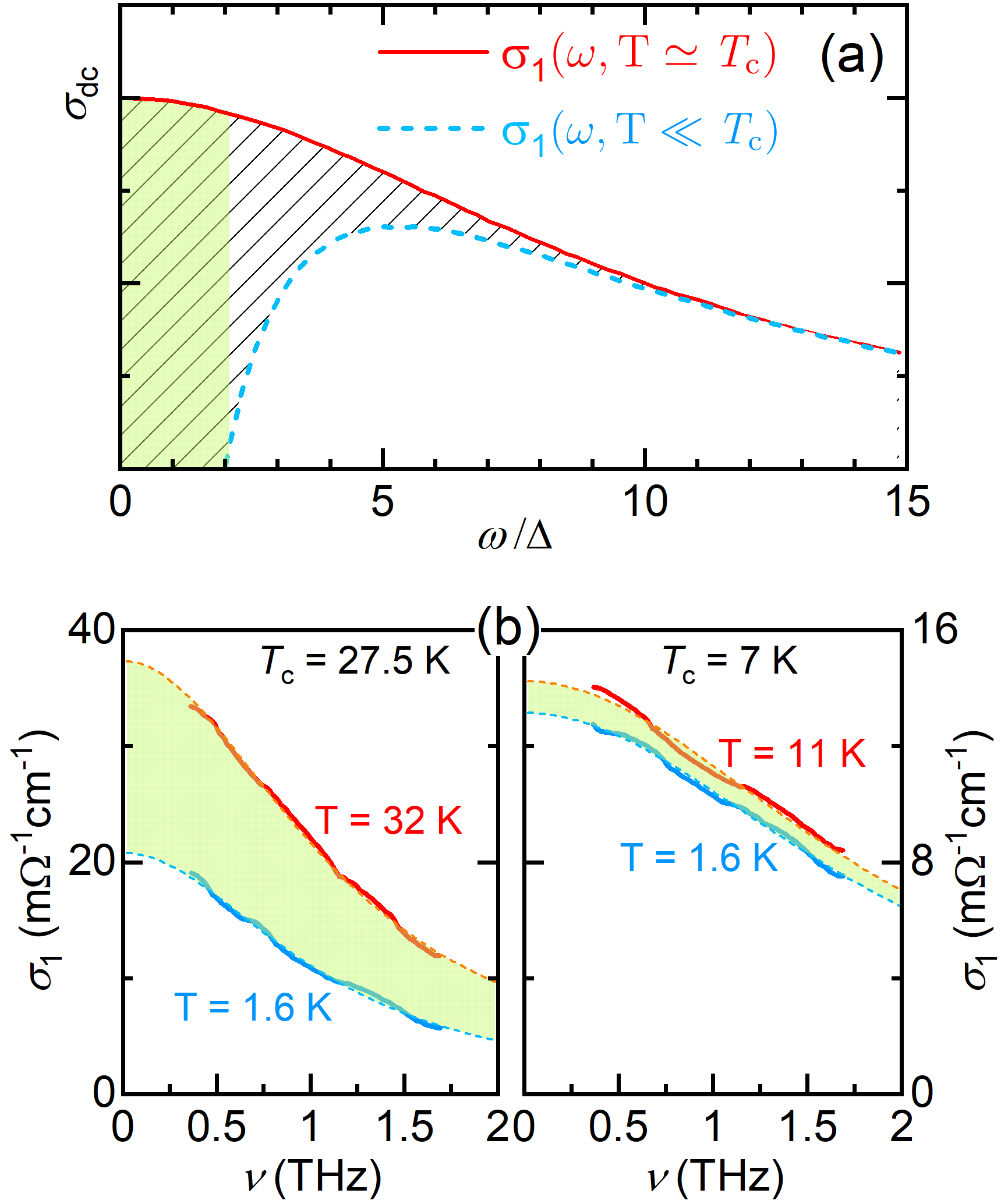}
    \caption{Optical conductivity above and below $T_c$ in (a) underdoped (schematic~\cite{Homes05}) and (b) very overdoped (as measured~\cite{Mahmood19}) regime. In the underdoped regime it is the itinerant (Fermi-liquid~\cite{NBarisic13,Chan14,Mirzaei13}) charges (\neff) of the normal state that collapse into the superconducting condensate~\cite{Homes04} (note that \nloc = 1, thus \rhoS~$\propto$~\neff). The missing spectral weight which is contained in the SC delta peak at $\omega = 0$ is schematically indicated with a hashed area. A rough estimate of the superfluid density, corresponding to the green area, is \rhoS~$\propto$~2$\Delta$ \sigdc, yielding the Homes law (which is usually presented on the log-log graph and thus not sensitive to details)~\cite{Homes04}. The Homes law holds well in the underdoped cuprates and simply states that \rhoS~$\propto$~\Tc \sigdc\ because, according to BCS, $\Delta$~$\propto$~\Tc~\cite{Homes05}. On the overdoped side, as obvious from (b) it is again the itinerant charges (\neff) which become superconducting, but the measured superfluid density~\cite{Mahmood19, Bozovic16} (corresponding to the green area between the red and blue curve) is proportional to the number of localized holes (\nloc), which is according to Eq.~\ref{eqn:rhoS}:~\rhoS~$\propto\neff    \cdot \nloc$~\cite{Pelc19}. Notably, both \nloc\ and \neff\  are obtained directly from the normal state. The reduction of the coherent spectral weight on the overdoped side is only partial simply because there are more itinerant particles
    than glue (localized holes).}
    \label{suppFig:Schematic_SpectWt}
\end{figure}

\subsection{Inspection of spectral weight transfers by eye}

Just by looking at Fig.~\ref{suppFig:LSCO_OptResp}, it becomes obvious, without any calculation, what the main effect of doping on the spectral weight distribution is. Namely, a spectral weight of exactly one hole per CuO$_2$ must be transferred from high energy (above the charge-transfer gap) to the FL peak. This simple conclusion follows from the fact that the spectral weight in the FL peak in the underdoped regime is equal to $p$~\cite{Padilla05} (so exactly 1 hole per CuO$_2$ is still localized~\cite{Pelc19,NBarisic15,NBarisic19}), while in the overdoped regime it essentially corresponds to $1+p$ (see Fig.~1 and related discussion). Thus, it is straightforward to conclude that it is the MIR feature which is responsible for this spectral weight transfer, as pointed out earlier \cite{Pelc19,NBarisic22}.

Finally, a rough examination of the doping evolution of the ``missing'' spectral weight associated with the superconductivity is also quite instructive (Fig.~\ref{suppFig:Schematic_SpectWt}). In the underdoped regime it is quite clear that the spectral weight of the FL peak is transferred to the superconducting condensate and thus \rhoS\ in this regime corresponds to \neff. This observation partially explains the Homes law \cite{Homes04,Homes05}. On the overdoped side, the situation is somewhat different. The spectral weight is similarly transferred from the FL component, but now in proportion to \nloc \cite{Pelc19,NBarisic22}. These two observations are captured by Eq.~\ref{eqn:rhoS} for \rhoS\, which is valid across the superconducting dome (with a small correction at the very edges due to the percolative nature of the superconductivity \cite{Pelc18,Popcevic18,Yu19a}).

\end{appendix}

\clearpage


%


\end{document}